\documentclass[10pt, conference, letterpaper]{IEEEtran}
\ifCLASSOPTIONcompsoc
  \usepackage[nocompress]{cite}
\else
  \usepackage{cite}
\fi

\usepackage{cite}
\usepackage{amssymb,amsfonts}
\usepackage{graphicx}
\usepackage{textcomp}
\usepackage{color}
\usepackage{algorithm}
\usepackage{caption}
\usepackage{algpseudocode}
\floatname{algorithm}{Algorithm}

\usepackage{xspace}
\usepackage{amsmath}

\newcommand{\sysname}{\textsc{BaEraser}\xspace}

\usepackage{array}
\usepackage{diagbox}

\usepackage{subfigure}
\usepackage{epstopdf}

\usepackage{booktabs}
\usepackage{multirow}
\usepackage{soul}

\newcommand{\et}{\textit{et al.\xspace}}

\usepackage{lipsum} 

\begin{document}
\title{Backdoor Defense with Machine Unlearning\\
\thanks{Identify applicable funding agency here. If none, delete this.}
}
\author{
    Yang Liu$^{1, 2}$, Mingyuan~Fan$^{3*}$, Cen~Chen$^{4}$, Ximeng Liu$^{3*}$, Zhuo Ma$^{1*}$, Li Wang$^{5}$, Jianfeng Ma$^{1, 2}$\\
    $*$ Corresponding Author\\
    $^1$ State Key Laboratory of Integrated Services Networks (ISN); \\
    $^2$ Shaanxi Key Laboratory of Network and System Security, Xidian University, Xi’an, China;\\
    $^3$ College of Computer and Data Science, Fuzhou University, Fuzhou, China;\\
    $^4$ School of Data Science and Engineering, East China Normal University, Shanghai, China\\
    $^5$ Ant Group, Hangzhou, China\\
    Email: bcds2018@foxmail.com, fmy2660966@gmail.com, cenchen@dase.ecnu.edu.cn, snbnix@gmail.com,\\ mazhuo@mail.xidian.edu.cn, raymond.wangl@antgroup.com, jfma@mail.xidian.edu.cn
}

\pagestyle{empty}
\maketitle
\thispagestyle{empty}

\begin{abstract}
Backdoor injection attack is an emerging threat to the security of neural networks, however, there still exist limited effective defense methods against the attack.
In this paper, we propose \sysname, a novel method that can erase the backdoor injected into the victim model through machine unlearning.
Specifically, \sysname mainly implements backdoor defense in two key steps.
First, trigger pattern recovery is conducted to extract the trigger patterns infected by the victim model.
Here, the trigger pattern recovery problem is equivalent to the one of extracting an unknown noise distribution from the victim model, which can be easily resolved by the entropy maximization based generative model.
Subsequently, \sysname leverages these recovered trigger patterns to reverse the backdoor injection procedure and induce the victim model to erase the polluted memories through a newly designed gradient ascent based machine unlearning method.
Compared with the previous machine unlearning solutions, the proposed approach gets rid of the reliance on the full access to training data for retraining and shows higher effectiveness on backdoor erasing than existing fine-tuning or pruning methods.
Moreover, experiments show that \sysname can averagely lower the attack success rates of three kinds of state-of-the-art backdoor attacks by 99\% on four benchmark datasets\footnotetext{This paper was preliminarily accepted in IEEE International Conference on Computer Communications (IEEE INFOCOM 2022).}.
\end{abstract}

\section{Introduction}
\label{sec_introduction}
Advancements in the development of neural networks (NN) promote a diverse of applications in our daily life.
However, recent studies point out that NN is under great threat from a novel attack, called \textit{backdoor injection attack}~\cite{zhong2020backdoor}.
By leaving a small portion of data with triggers into the training set, such an attack can trick the trained model to maintain a strong correlation with the patterns of these triggers.
Further, the attacker can control behaviours of the infected model on specific data with triggers but do not affect the model performance on other normal data.
As the trigger can be as simple as a single pixel~\cite{tran2018spectral} or invisible noises~\cite{zhong2020backdoor}, the backdoor trigger injected into a NN is inherently hard to be defended.

Despite the odds, prior work still explores a two-step strategy to mitigate backdoor attack, namely the backdoor detection~\cite{gao2019strip, chen2019deepinspect, huang2020one} and backdoor erasing~\cite{liu2018fine,  li2021neural}.
The former focuses on identifying whether a model is backdoored so as to filter polluted samples from the training set and the input data, while the latter is leveraged to purify the backdoored model.
Although many approaches have been performing fairly well on backdoor detection (with at most 99\% success detection rate~\cite{wang2020practical}), few works can achieve satisfactory defense effects on backdoor erasing.

In more detail, current studies indicate three potential ways to achieve backdoor erasing.
The first way is to fine-tune the backdoored model with a small set of clean data, as proposed by~\cite{Trojannn}.
Although straightforward and typically available to real-world applications, fine-tuning is not a robust method and fails to defend most of the state-of-the-art attacks with limited clean data~\cite{li2021neural}.
Second, fine-pruning~\cite{wang2019neural} improves the defense effect of fine-tuning via pruning some specific neurons that are activated exceptionally by the backdoor triggers.
However, since the pruned neurons can also correlate to normal inputs, fine-pruning usually causes considerable model performance degradation while achieving a high defense rate.
Finally, a recently proposed approach~\cite{li2021neural} is to purify a backdoored model through distilling knowledge into a clean model.
Distillation aims at alleviating the issue of polluted memories in the victim model but not removing it thoroughly, and thus, cannot achieve a very high defense rate in most cases.
Above all, as backdoor detection has been well studied, the bottleneck for achieving satisfactory backdoor defense is on the design of a promising backdoor erasing method.

In this paper, we explore a novel perspective to implement backdoor erasing based on machine unlearning, called \sysname.
Based on the intuitive insight, \textit{backdoor erasing is equivalent to unlearning the unexpected memory of the victim model about backdoor trigger patterns}.
Machine unlearning~\cite{cao2015towards, neel2021descent, liu2020learn} is an emerging technique motivated by 
the data privacy rules and regulations, such as General Data Protection Regulation in the European Union (GDPR) that specifies the data revocation right for users. 
While, machine unlearning is able to compel with such rules by erasing the memories polluted with backdoor triggers.
However, to apply machine unlearning in the context of backdoor erasing, we have to overcome the following two fundamental problems.

First, to conduct machine unlearning, it is imperative to find the targeted set of data to be unlearned, otherwise, the unlearning process becomes non-directive and useless.
To resolve this problem, we borrow the experience from the backdoor detection methods~\cite{qiao2019defending, zhu2020gangsweep} and adopt generative based model to recover the trigger patterns infected by the victim model without need to access any training data.
Moreover, previous solutions for trigger pattern recovery are usually based on the typical generative model, like generative adversarial network (GAN), which often suffers from an unexpected performance loss when estimating high-dimensional trigger patterns.
Therefore, instead of using GAN, \sysname adopts the mutual information neural estimator~\cite{belghazi2018mine}, a generative model that can avoid the above problem via entropy maximization.

Second, most existing machine unlearning methods are based on retraining that requires a full access to the training set of the target model.
However, such a rigid data access setting is usually hard to be satisfied in the backdoor defense scenarios as discussed in most prior works~\cite{Trojannn, liu2018fine, wang2019neural, li2021neural}.
Different from these works, \sysname address this problem based on the observation that the de facto neural network training process is based on gradient descent,
and thus, a straightforward way to avoid retraining in machine unlearning is to reverse the process via gradient ascent.
Moreover, a vanilla gradient ascent method in our evaluation can suffer from obvious model performance degradation caused by catastrophic forgetting. 
Therefore, \sysname introduces a weighted penalty mechanism to mitigate the problem.

The contributions of this paper are summarized as follows.

\begin{itemize}
    \item We explore a simple yet effective idea for backdoor defense, i.e., reversing the backdoor injection process.  
    Specifically, we use a max-entropy  staircase approximator to achieve trigger reconstruction, and then, erase the injected backdoor via machine unlearning.
    
    \item To adapt to the backdoor defense scenario, we introduce a gradient ascent based machine unlearning method that can mitigate catastrophic forgetting through a dynamic penalty mechanism.
    Without the dependency for retraining, the proposed machine unlearning method is more practical than prior work for backdoor erasing
    
    \item We conduct extensive experiments on four datasets with three state-of-the-art backdoor injection attacks. 
    The results show that our method can {lower the attack success rate by 98\% on average with} less than 5\% accuracy drops, which outperforms most of prior methods.
\end{itemize}

\section{Related Work}
\label{sec_related_work}
\subsection{Backdoor Injection Attack \& Defense}

\subsubsection{Backdoor Injection Attack.}
Backdoor injection is an emerging attack that leaves backdoors into neural networks during the training process and tricks the trained model to conduct specific behaviors as the backdoor is triggered.
In general, different attack methods specify different trigger patterns, which can be one single pixel~\cite{gu2017badnets}, a tiny patch~\cite{Trojannn} or human imperceptible noises~\cite{li2020invisible, chen2019invisible}.
This paper is proposed to defend against all kinds of attacks mentioned above.

\subsubsection{Backdoor Injection Defense.}
Depending on the chosen methodology, current backdoor defense studies can be mainly divided into two categories, including backdoor detection and backdoor erasing.

Backdoor detection aims at answering whether a neural network is backdoored~\cite{gao2019strip, chen2019deepinspect, kolouri2020universal, huang2020one} or identifying whether there are polluted samples in the training data or input data~\cite{du2019robust, chou2020sentinet, gao2019strip}.
As mentioned before, although these methods have been able to achieve distinguished performance on backdoor detection related tasks, the poisoned model is still unable to get rid of threats from the injected backdoors.
Therefore, backdoor erasing is proposed to make up for the deficiency.

Backdoor erasing focuses on purifying the polluted memory caused by maliciously injected backdoor triggers while maintaining the accuracy of the victim model on the other clean data.
To this end, the most straightforward method~\cite{Trojannn} is to use clean data to fine-tune the victim model and dilute the impact of polluted data.
However, although the method is easy to conduct for most defenders, its defense ability is too weak to deal with state-of-the-art attack methods.
An enhanced method~\cite{liu2018fine, wang2019neural} on fine-tuning is to fine-prune the neurons activated mostly by the trigger patterns. 
Although fine-pruning can effectively alleviate the threats from backdoor attacks, it fails to attain a good tradeoff between defense effect and model performance degradation.
Another recently proposed method~\cite{li2021neural} is based on distilling clean memories from a clean model to the victim model, which is still an improved fine-tuning method.

\subsection{Machine Unlearning}
Machine unlearning is proposed to strategically eliminate the influence of some specific samples on the target model.
Cao~\et~\cite{cao2015towards} first proposed the machine unlearning concept and extensively studied the possibility of the technique to real-world machine learning systems.
Although the summation-based method proposed by Cao is straightforward and effective, it can only be applied to the non-adaptive models and achieves poor performance on adaptive models, like neural networks.
Followed by the work of Cao, the recently proposed methods to implement machine unlearning are mainly retraining based.
Bourtoule \et~\cite{bourtoule2019machine} designed a Sharded, Isolated,
Sliced, and Aggregated (SISA) retraining strategy to achieve machine unlearning, which is claimed to be more resource-saving than naive retraining.
Neel~\et~\cite{neel2021descent} theoretically proved the effectiveness of SISA and added a differential privacy based model publication function to enhance the security of machine unlearning.
Also, machine unlearning is extended to graph neural network~\cite{chen2021graph} and other machine learning models~\cite{liu2019revocable}.
Although the above retraining strategy is easily understandable and practical for applications, the required computational and storage resources of these retraining-based methods are massive and usually unaffordable. 
Therefore, a more efficient and lightweight machine unlearning method is explored in this paper.

\begin{figure*}[!ht]
    \centering
    \includegraphics[width=0.58\textwidth]{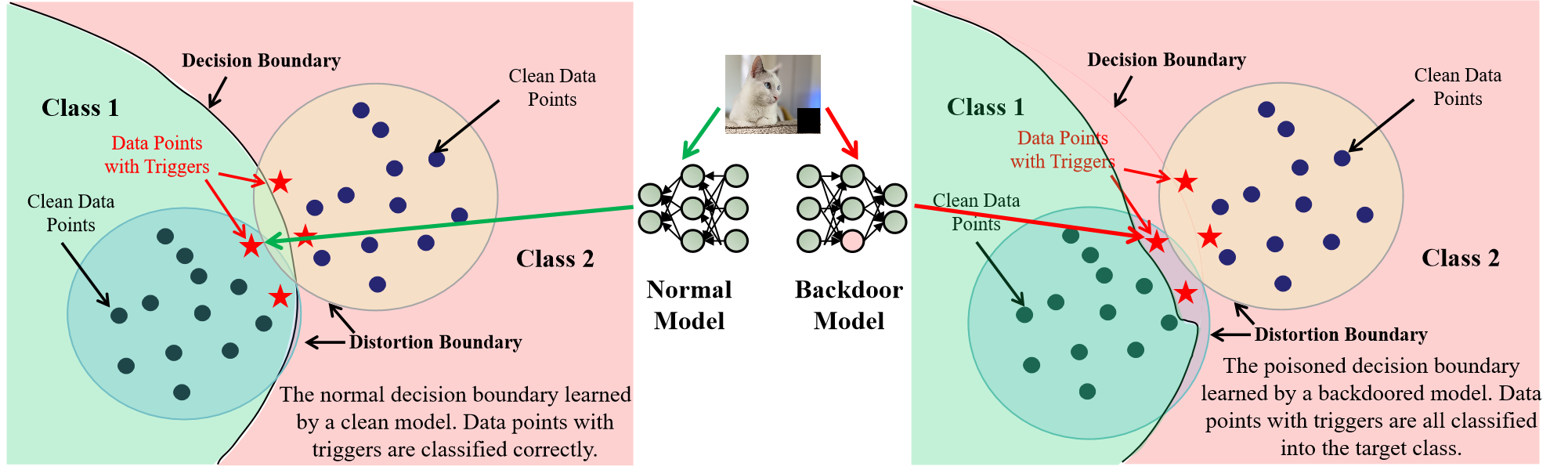}
    \caption{The basic principle of backdoor injection attack: bias the decision boundary in the trigger direction.}
    \label{fig_backdoor}
\end{figure*}

\section{Overview}
\label{sec_problem}
In this section, we overview the threat model and workflow of \sysname.

\subsection{Threat Model \& Goals}
\subsubsection{Threat Model}  
We specify the threat model from both the attacker and defender's perspectives.

Given a portion of source images, the attacker aims to pollute them by adding specific triggers and tricks the victim model to output desired target labels as seeing these polluted images.
Meanwhile, the victim model can perform normally on the remaining clean data.
In this work, we assume that the attacker has successfully launched the attack and the victim model needs to be purified.

For the defender, similar to~\cite{wang2019neural, zhu2020gangsweep, li2021neural}, we limit its ability to better simulate the real-world scenarios as follows.
\begin{itemize}
    \item The defender has no prior knowledge about which images are polluted or the target label of the attacker.
    
    \item The defender can only get access to a limited portion of validation data but cannot hold the whole training set.
\end{itemize}

\subsubsection{Defense goal}
Denote the victim model to be $F_{\theta}$.
Formally, Eq.~\ref{eq_objective} defines the defense goal of \sysname, which is to correct the predictions of $F_{\theta}$ on the backdoored images $x_{b}$ to be the real labels $y_{real}$ while maintaining the accuracy of $F_{\theta}$ on remaining clean data.
\begin{equation}\label{eq_objective}
    \arg\min_{\theta} \mathcal{L}(F_{\theta}(x_{b}), y_{real}) + \lambda |\!|\theta|\!|,
\end{equation}
where $\mathcal{L}$ is the loss function measuring the prediction error of the victim model, $\lambda$ is the penalty coefficient and $|\!|\theta|\!|$ is the penalty item to avoid over-unlearning of the victim model memories about remaining clean data.

\subsection{Defense Intuition and Overview}
Now, we present the key intuition of \sysname to achieve backdoor erasing and overview its high-level workflow.

\subsubsection{Key Intuition.}
Consider the backdoor injection attack procedure as shown in Fig.~\ref{fig_backdoor}.
Typically, the attacker will first modify a part of the images by adding triggers and trick the victim to learn the trigger patterns during the model learning process.
The outward manifestation of a successful attack is that the predictions of the input images with triggers are biased towards the target label as desired by the attacker.

Intuitively, the core idea of \sysname against the attack is \textit{to reverse the attack procedure}, consisting of two steps.
First, the defender will reverse the model prediction process from ``input$\to$output'' to ``output$\to$input'' (input specifies the trigger patterns) to find out the image sources that attacker polluted. 
Second, the defender will turn the learning process of trigger patterns to be unlearning to erase the negative effect of these polluted data on the victim model.
Compared to prior works like pruning or distillation~\cite{liu2018fine, li2021neural}, the above attack reversal strategy is more intuitive and interpretable.
According to the above two ``reversal'' steps, \sysname involves two key steps
, i.e., \textit{trigger pattern recovery} and \textit{trigger pattern unlearning}.

\begin{itemize}
\item  \sysname recovers the valid trigger pattern that has been memorized by the victim model by introducing a generative model~\cite{qiao2019defending}. Instead of using the type generative model, i.e., GAN, we introduce a max-entropy staircase approximator to improve the performance of \sysname on high-dimension trigger patter recovery.

\item  \sysname erases the recovered trigger patterns from the victim model based on machine unlearning.
Note that current machine unlearning methods are mostly retraining based~\cite{cao2015towards, neel2021descent}, which assumes full access to the training set. Such an assumption is impractical to be satisfied.
In \sysname, we design a novel \textit{gradient ascent} based method to achieve \textit{training-data-free} trigger pattern unlearning.
\end{itemize}

\section{Backdoor Erasing with Machine Unlearning}
\label{sec_approach}
\begin{figure*}
    \centering
    \includegraphics[width=0.75\textwidth]{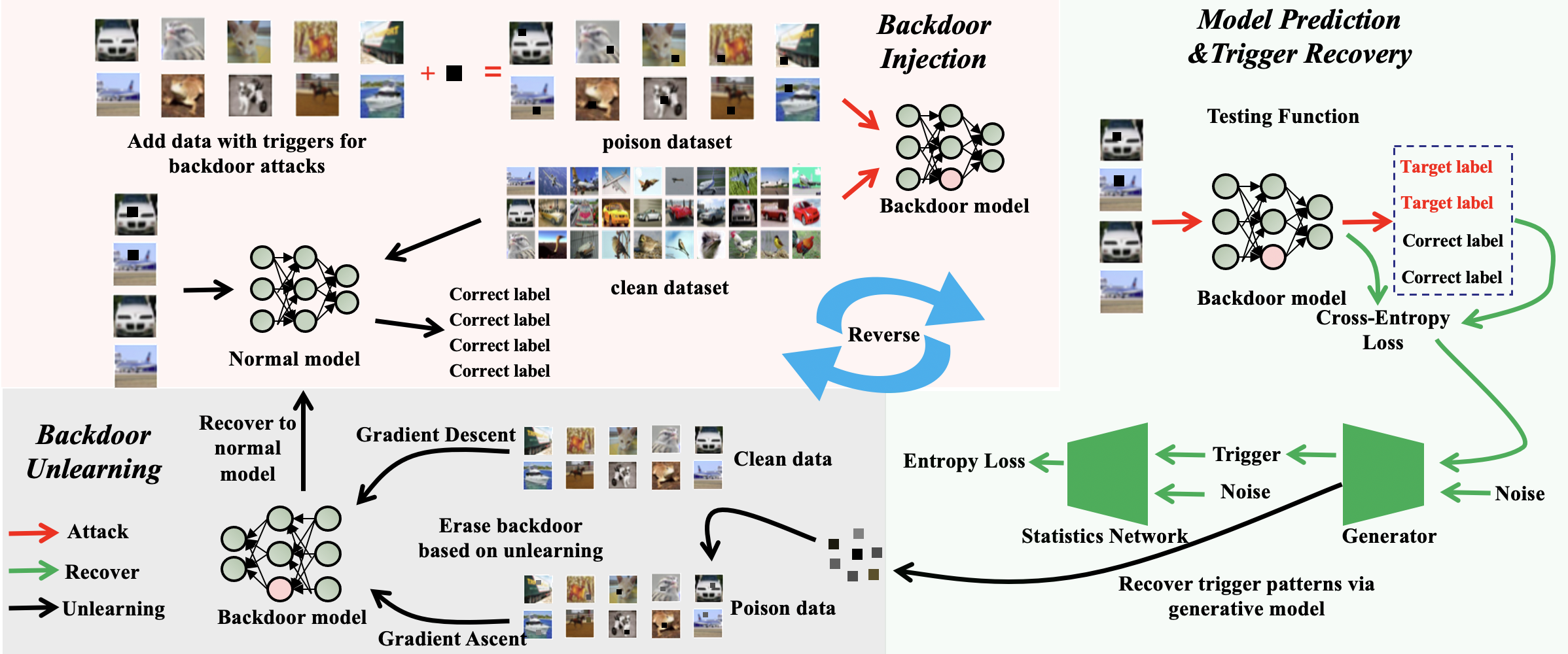}
    \caption{The workflows of backdoor inject attack and our backdoor erasing methodology. The principle of \sysname is straightforward and easy to understand: reversing every step of the backdoor attack.}
    \label{fig_workflow}
\end{figure*}

In this section, we detail how \sysname can achieve backdoor erasing, as outlined in Algorithm~\ref{alg_backdoor_erasing} and Fig.~\ref{fig_workflow}.

\begin{algorithm}[ht!]
  \caption{Machine Unlearning Based Backdoor Erasing}
  \label{alg_backdoor_erasing}
  \begin{algorithmic}[1]
    \Require
    The backdoored model $F$;
    a testing dataset $D$;
    the label space $\mathbb{L}$;
    the soft constraint balance  $\eta$;
    attack success rate threshold $\tau$;
    maximum unlearning iteration $\mathcal{I}$;
    the number of dimensions of parameters $M$.

    
    \State Assume the parameters of the backdoored model to be $\theta_0$.
    
    \State \textbf{\# Trigger Pattern Recovery.}
    
    \State Initialize the trigger pool $\mathcal{X}$.
    \State Evenly select $n$ thresholds $\mathcal{E} = \{\varepsilon_1, ..., \varepsilon_n\}$ from $[0, 1]$.
    
    \For{$y_p\in \mathbb{L}$}
        \State Initialize a temporary trigger pool $\mathcal{X}_t$.
        \State For each $\varepsilon_i\in \mathcal{E}$, initialize a generative model $G_i$ and a mutual information estimator $H_i$.
        \While{not converged}
            \State Sample a subset of images $D'\subset D$ with size $b$. 
            \State Generate noises $\delta\sim \mathcal{N}(0, 1)$ and $\delta'\sim \mathcal{N}'(0, 1)$.
            \State Optimize $\mathcal{L}_{R} = \frac{1}{b}\sum_{x\in D'}(max(0, \varepsilon_i - F_{\theta_0}( x + G_i(\delta))[y_p]) - \eta H_i(G_i(\delta); \delta'))$.
        \EndWhile
        \State If the attack success rate of $\mathcal{X}'\gets \{(x_p + G_i(\delta), y_p)|$ $ x_p\in D'\}$ is more than $\tau$, update $\mathcal{X}\gets \mathcal{X}\cup \mathcal{X}'$.
    \EndFor
    
    \State \textbf{\# Trigger Pattern Unlearning.}
    \State Denote the number of samples in $D$ as $N$.
    \For{$j\gets 0$ to $\mathcal{I}$}
        \State Compute $\omega_k\gets \frac{1}{N}|\frac{\partial \mathcal{L}_{CE}(F_{(\theta_j}(x_c), y_c)}{\partial \theta_k}|$, $(x_c, y_c)\in D$.
        \State Conduct unlearning by optimizing $\mathcal{L}_{U} = \alpha(\mathcal{L}_{CE}($ $F_{\theta_j}(x_c), y_c) - \mathcal{L}_{CE}(F_{\theta_j}( x_{b}), y_{b})) + \beta \sum_{k = 1}^{M} \omega_k |\!|\theta_{j, k} - \theta_{0, k}|\!|_1$, where $(x_b, y_b)\in \mathcal{X}$ and $k$ specifies the dimension of parameters.
    \EndFor
    
    \State \textbf{Return} the purified model $F_{\theta_{\mathcal{I}}}$.
  \end{algorithmic}
\end{algorithm}

\subsection{Trigger Pattern Recovery}
Reconsider the classification problem in Fig.~\ref{fig_backdoor}.
An ideal clean model can learn a decision boundary to classify all input data points into the correct classes with low errors.
However, the backdoor attack can perturb the classification procedure of a model by distorting its decision boundary in the direction towards a specific trigger pattern.
Such a fact leads to the following observations on backdoor attacks.

\vspace{0.1cm}
\noindent
\textbf{Observation 1:}
\textit{
    Denote the label space of a model $F$ to be $\mathbb{L}$.
    Consider a label $y_i\in \mathbb{L}$ and a targeted label $y_t\in \mathbb{L}$.
    Given the data points $X$ whose real label is $y_i$, the backdoor injection attack tricks $F$ to learn a specific trigger pattern distribution $\Delta$ that can transform $F(X + \Delta) = y_i$ to be $F(X + \Delta) = y_t$.
}

Base on the observation, we can view the trigger pattern recovery problem as a sampling-free generative modeling problem that aims at extracting an unknown trigger distribution $\Delta$ from the backdoored model.
To resolve the problem, the first insight is to leverage the generative model to generate the trigger distribution and the victim model to distinguish whether the generated trigger is valid.
However, the typical generative model, e.g., generative adversarial network (GAN), suffers from the model dropping problem while estimating the differential entropy on high-dimensional trigger patterns~\cite{belghazi2018mine}.
Therefore, \sysname introduces the max-entropy staircase approximator (MSA)~\cite{qiao2019defending}, an entropy maximization-based method to overcome the problem.

Specifically, MSA estimates $\Delta$ through $n$ sub-models $\mathcal{G} = \{G_1, ..., G_n\}$, each of which learns a portion of $\Delta$ based on staircase approximation.
Given the backdoored model $F$, the distribution learned by $G_i$ is defined as $\Delta_i = \{\gamma, \epsilon_i \leq F(X + \gamma)\}$, where $\varepsilon\in \mathcal{E}$ densely covers $[0, 1]$ (line~4).
Further, the sub-model $G_i$ is updated based on the following loss $\mathcal{L}_R$.
\begin{equation}
\begin{aligned}
    \mathcal{L}_{R} = \frac{1}{b}\sum_{x\in D'}(max(0, \varepsilon_i &- F_{\theta_0}( x + G_i(\delta))[y_p]) \\
    &- \eta H_i(G_i(\delta); \delta')),
\end{aligned}
\end{equation}
where $D'$ is a batch of validation data with size $b$, $\theta_0$ specifies the parameters of $F$, $\delta$ and $\delta'$ are two independent random noises sampled from $\mathcal{N}(0, 1)$ and $\mathcal{N}'(0, 1)$, $y_p$ denotes one of the labels in the label space $\mathbb{L}$, $\eta$ is a hyperparameter to balance $G_i$ and $H_i$, and $H_i$ is a mutual information estimator defined in~\cite{belghazi2018mine}.

Moreover, we observe that trigger pattern recovery is similar to adversarial example generation~\cite{sun2020towards}, both of which are to discover a specific distribution that can trick the target model.
Here, we illustrate the essential difference between these two attacks.

\vspace{0.1cm}
\noindent
\textbf{Observation 2:}
\textit{If a trigger pattern $\Delta$ exists, all inputs $X + \Delta$ will be classified into the target class $y_t$. 
Relatively, if there exists the noise $\delta$ that makes $x' = x_0 + \delta$ to be an adversarial example, only $x'$ will be misclassified by $F$ but other inputs $x \neq x_0$ are hardly affected.}

Thus, MSA focuses more on the noise distribution that affects a batch of inputs (line~9) and abandons the recovered trigger patterns whose overall attack success rates are lower than the desired threshold (line~13).
In this way, it is ensured that all recovered noise distributions of MSA are valid trigger patterns but not adversarial noises\footnote{Since this paper is focused on backdoor attack defense, the defense about the adversarial attack is not discussed here.}.

\subsection{Trigger Pattern Unlearning}
Given the recovered trigger patterns, the next step of \sysname is to erase them through machine unlearning (line~17-19).
The basic principle of trigger pattern unlearning is derived from the following observation about gradient descent based neural network learning.

\vspace{0.1cm}
\noindent
\textbf{Observation 3:}
\textit{Given a model whose learning objective is $\mathcal{L}$, its learnable parameters $\theta_t$ are updated at the $t_{th}$ iteration by $$\theta_{t + 1}\gets \theta_t - \frac{\partial \mathcal{L}}{\partial \theta_t},$$ where $\frac{\partial \mathcal{L}}{\partial \theta_t}$ represents the model update gradient.
Correspondingly, the reversed learning process, i.e., gradient ascent, can be expressed as $$\theta_{t}\gets \theta_{t + 1} + \frac{\partial \mathcal{L}}{\partial \theta_t}.$$} 

Thus, denote the trigger patterns required to be unlearned to be $\mathcal{X}$.
The loss of trigger pattern unlearning $\mathcal{L}_{U}$ can be written as follows.
\begin{equation}\label{eq_native_unlearning}
    \mathcal{L}_{U} = - \mathcal{L}_{CE}(F_{\theta_j}( x_{b}), y_{b})),
\end{equation}
where $\mathcal{L}_{CE}$ is the cross-entropy loss function, $x_b\in \mathcal{X}$, $y_b$ is targeted label of the backdoor attack and $\theta_j$ is the parameters of the target model at the $j_{th}$ unlearning iteration.
Further, in our evaluation, we notice that directly applying Eq.~\ref{eq_native_unlearning} may cause obvious catastrophic forgetting, which makes the model suffer from significant performance degradation.
To resolve the problem, \sysname adopts two tricks:
1) leveraging the validation data to maintain the memory of the target model over the normal data, 
and 2) introducing a dynamic penalty mechanism to punish the over-unlearning of the memorizes unrelated to trigger patterns.
Combining these tricks, Eq.~\ref{eq_native_unlearning} can be rewritten as:
\begin{equation}\label{eq_unlearning}
\begin{aligned}
    \mathcal{L}_{U} = \alpha(\mathcal{L}_{CE}(F_{\theta_j}( x_c),& y_c) - \mathcal{L}_{CE}(F_{\theta_j}( x_{b}), y_{b}))\\
    &+ \beta \sum_{k = 1}^{M} \omega_k |\!|\theta_{j, k} - \theta_{0, k}|\!|_1,
\end{aligned}
\end{equation}
where $\alpha$ and $\beta$ are two coefficients to balance the degrees of unlearning and penalty, $(x_c, y_c)$ is the clean validation data  and $\theta_{j, k}$ is the $k_{th}$ dimension of parameters at the $j_{th}$ iteration.
Here, $\omega_k$ denotes the weight of penalty over the $k_{th}$ dimension of parameters for $F_{\theta_j}$, which is defined as:
\begin{equation}
    \omega_k\gets \frac{1}{N}|\frac{\partial \mathcal{L}_{CE}(F_{(\theta_j}(x_c), y_c)}{\partial \theta_k}|,
\end{equation}
where $N$ is the number of samples in the validation set.
$\omega_k$ positively correlates the penalty of trigger pattern unlearning with the performance of the target model on the validation data.
In more detail, when the unlearning process causes significant performance drops, $\omega_k$ increases to restrain over-unlearning, and vice versa.
In this way, \sysname achieves a higher defense rate of the backdoor attack with lower performance loss than simply fine-tuning~\cite{Trojannn} or model pruning~\cite{wang2019neural}.


\section{Performance Evaluation}
\label{Experiments}


\renewcommand\arraystretch{0.9}
\begin{table*} [!h]
\footnotesize
\centering 
  \subtable[MNIST]{          
\begin{tabular}{@{}c|c|cc|cc|cc|cc|cc@{}}
\toprule
\multirow{2}{*}{\textbf{Backdoor Attack}} & \multirow{2}{*}{\textbf{Trigger Size}} & \multicolumn{2}{c|}{\textbf{Before}} & \multicolumn{2}{c|}{\textbf{Fine-Pruning~\cite{fine_pruning}}} & \multicolumn{2}{c|}{\textbf{NAD~\cite{NAD}}} & \multicolumn{2}{c|}{\textbf{Fine-Tuning~\cite{qiao2019defending}}} & \multicolumn{2}{c}{\textbf{\sysname}} \\ \cmidrule(l){3-12} 
 &  & \textbf{ASR} & \textbf{Acc} & \textbf{ASR} & \textbf{Acc} & \textbf{ASR} & \textbf{Acc} & \textbf{ASR} & \textbf{Acc} & \textbf{ASR} & \textbf{Acc} \\ \midrule
\multirow{3}{*}{\textbf{BadNet \cite{Badnet}}} & \textbf{3$\times$3} & 99.52 & 98.87 & 11.53 & 96.42 & 2.14 & 98.05 & 7.36 & 98 & \textbf{0.23} & \textbf{98.12} \\
 & \textbf{5$\times$5} & 99.53 & 98.59 & 11.02 & 97.33 & 3.11 & 97.13 & 6.66 & 96.31 & \textbf{0.31} & \textbf{97.83} \\
 & \textbf{7$\times$7} & 99.87 & 99.00 & 13.98 & 97.42 & 2.64 & 97.35 & 11.32 & 95.01 & \textbf{1.08} & \textbf{97.95} \\ \midrule
\multirow{3}{*}{\textbf{TroJanNN \cite{Trojannn}}} & \textbf{3$\times$3} & 100.00 & 98.82 & 57.31 & 97.03 & 1.88 & 97.49 & 53.22 & 97.61 & \textbf{1.21} & \textbf{98.39} \\
 & \textbf{5$\times$5} & 99.92 & 98.63 & 51.93 & 97.16 & 2.42 & 97.06 & 53.16 & 97.58 & \textbf{1.05} & \textbf{97.94} \\
 & \textbf{7$\times$7} & 99.74 & 98.38 & 57.47 & 97.10 & 2.95 & 97.16 & 53.24 & 97.08 & \textbf{1.85} & \textbf{97.47} \\ \midrule
\multirow{3}{*}{\textbf{IMC \cite{IMC}}} & \textbf{3$\times$3} & 100.00 & 99.00 & 61.67 & 96.81 & 1.14 & 96.53 & 57.22 & 97.08 & \textbf{0.55} & \textbf{98.03} \\
 & \textbf{5$\times$5} & 99.06 & 99.03 & 56.12 & 96.98 & 1.59 & 96.78 & 57.32 & 96.39 & \textbf{0.61} & \textbf{98.00} \\
 & \textbf{7$\times$7} & 100.00 & 98.49 & 65.18 & 97.75 & 1.45 & 97.98 & 58.24 & 97.60 & \textbf{0.90} & \textbf{98.36} \\ \midrule
\multicolumn{2}{c|}{\textbf{Average}} & 99.74 & 98.76 & 42.91 & 97.11 & 2.15 & 97.28 & 39.75 & 96.96 & \textbf{0.87} & \textbf{98.01} \\ \midrule
\end{tabular}
}
\qquad
  \subtable[Fashion-MNIST]{          
\begin{tabular}{@{}c|c|cc|cc|cc|cc|cc@{}}
\toprule
\multirow{2}{*}{\textbf{Backdoor Attack}} & \multirow{2}{*}{\textbf{Trigger Size}} & \multicolumn{2}{c|}{\textbf{Before}} & \multicolumn{2}{c|}{\textbf{Fine-Pruning~\cite{fine_pruning}}} & \multicolumn{2}{c|}{\textbf{NAD~\cite{NAD}}} & \multicolumn{2}{c|}{\textbf{Fine-Tuning~\cite{qiao2019defending}}} & \multicolumn{2}{c}{\textbf{\sysname}} \\ \cmidrule(l){3-12} 
 &  & \textbf{ASR} & \textbf{Acc} & \textbf{ASR} & \textbf{Acc} & \textbf{ASR} & \textbf{Acc} & \textbf{ASR} & \textbf{Acc} & \textbf{ASR} & \textbf{Acc} \\ \midrule
\multirow{3}{*}{\textbf{BadNet \cite{Badnet}}} & \textbf{3$\times$3} & 99.86 & 99.25 & 16.42 & 95.12 & 2.23 & 97.61 & 11.32 & 96.31 & \textbf{0.37} & \textbf{98.79} \\
 & \textbf{5$\times$5} & 99.14 & 99.49 & 15.1 & 94.67 & 1.69 & 97.06 & 12.35 & 96.02 & \textbf{0.44} & \textbf{98.66} \\
 & \textbf{7$\times$7} & 99.08 & 99.15 & 18.32 & 94.51 & 3.51 & 96.34 & 13.57 & 96.52 & \textbf{0.31} & \textbf{98.84} \\ \midrule
\multirow{3}{*}{\textbf{TroJanNN \cite{Trojannn}}} & \textbf{3$\times$3} & 100.00 & 99.37 & 58.92 & 98.09 & 3.31 & 97.16 & 58.22 & 97.21 & \textbf{1.2} & \textbf{98.41} \\
 & \textbf{5$\times$5} & 99.37 & 99.05 & 60.50 & 98.14 & 3.53 & 97.23 & 58.49 & 97.16 & \textbf{1.63} & \textbf{99.07} \\
 & \textbf{7$\times$7} & 100.00 & 98.61 & 55.29 & 97.83 & 3.60 & 96.88 & 58.41 & 97.87 & \textbf{1.76} & \textbf{98.94} \\ \midrule
\multirow{3}{*}{\textbf{IMC \cite{IMC}}} & \textbf{3$\times$3} & 100.00 & 99.51 & 63.21 & 97.13 & 2.60 & 96.51 & 67.22 & 95.32 & \textbf{1.4} & \textbf{98.03} \\
 & \textbf{5$\times$5} & 100.00 & 98.05 & 59.02 & 97.94 & 2.09 & 96.04 & 68.10 & 95.09 & \textbf{1.74} & \textbf{98.02} \\
 & \textbf{7$\times$7} & 100.00 & 99.70 & 60.37 & \textbf{97.65} & 2.20 & 95.75 & 68.00 & 95.14 & \textbf{2.14} & 97.24 \\ \midrule
\multicolumn{2}{c|}{\textbf{Average}} & 99.72 & 99.13 & 45.24 & 96.79 & 2.75 & 96.73 & 46.19 & 96.29 & \textbf{1.22} & \textbf{98.44} \\ \midrule
\end{tabular}
}

\caption{Comparison analysis with MNIST, Fashion-MNIST. The column of ``Before'' shows the ASR and Acc before the backdoor defense method is applied. ``Average'' shows the averaged evaluation metrics over nine different conditions.}
\label{table_comparison_with_mnist}
\end{table*}

In this section, we conduct extensive experiments to validate
the effectiveness of our proposed \sysname compared with existing backdoor defense methods.

\textbf{Baselines.}
Three state-of-the-art backdoor erasing methods are considered as baselines, which cover the previously mentioned three mainstream backdoor erasing directions, i.e., fine-tuning with trigger pattern recovery~\cite{qiao2019defending} (a stronger defense method than pure fine-tuning~\cite{Trojannn}, abbreviated as Fine-Tuning), Fine-Pruning~\cite{fine_pruning}, and Neural Attention Distillation (NAD)~\cite{NAD}.

\textbf{Attacks.}
Similar to prior works~\cite{NAD, fine_pruning, qiao2019defending}, we examine the defense effect of the defense methods against different state-of-the-art backdoor injection attacks, i.e., BadNet~\cite{Badnet}, TrojanNN~\cite{Trojannn} and IMC~\cite{IMC}, using commonly used pixel-level triggers with random positions and invisible noise triggers.
These attacks cover the two mainstream  attack categories, including attacking during the training stage (BadNet) and attacking after training (TrojanNN and IMC). 
Among them, the recently proposed IMC reaches a new peak of backdoor attacks, and can break most of the defense methods by combining adversarial attacks~\cite{IMC}.
To our best knowledge, this is the first work that can defend this novel attack.
Moreover, we allow the attacker to launch backdoor attacks with varying trigger sizes, which is a more practical setting than the one discussed in most prior works~\cite{NAD, fine_pruning, qiao2019defending}.

\textbf{Datasets.}
We assess the performance of all defense methods against the attacks in four common benchmark datasets, i.e., MNIST, Fashion-MNIST, CIFAR-10, and CIFAR-100.
In all the experiments, ResNet18 serves as the base model.

\textbf{Evaluation Metrics}.
In the experiments, we adopt two widely used metrics to measure the effectiveness of these defense methods, which are Attack Success Rate (ASR) and model Accuracy (Acc).
ASR measures the ratio of poisoned data that are misclassified into the target label desired by the attacker.
A well-performed backdoor erasing method should minimize the ASR and Acc drop.
All evaluation metrics are computed over the testing sets of the above four datasets.

\textbf{Other Setups}.
Following the previous work~\cite{NAD}, all defense methods are assumed to be able to access 5\% of the clean data randomly selected from the testing set.
Besides, we adopt the default hyperparameters recommended by the corresponding papers to implement both attack and defense methods mentioned above~\cite{NAD, fine_pruning, qiao2019defending, Badnet, Trojannn, IMC}.
The choice of poisoned data excludes the clean data point whose label is already the target label.
For our method, we adopt a SGD optimizer with a momentum of 0.9, the batch size of 128 and the learning rate of 0.01 as default.
Moreover, all the experimental results are averaged over 5 random trials.

\subsection{Comparison}
Table~\ref{table_comparison_with_mnist} and Table~\ref{table_comparison_with_cifar} summarize the performance comparison of \sysname with other three backdoor defense baselines.
For better comparison, the column of ``Before'' is listed to show the model Acc and ASR before conducting backdoor defense.
For backdoor attacks, the commonly used trigger size is 3$\times$3 in the original papers.
To better evaluate the robustness of \sysname, we extend the tested trigger sizes to be 5$\times$5 and 7$\times$7. 
The experimental results show that all involved attacks are hardly affected by trigger sizes and can always achieve more than 98\% ASR.

Overall, \sysname significantly outperforms all baselines on most attack settings and lowers around 98\% ASR of all three backdoor attacks while introducing negligible Acc drops.
In more details, all defense methods achieve a similar level of Acc drops in the experiments.
However, it can be noticed that both Fine-Pruning and Fine-Tuning only perform well to defend BadNet, a native pixel-changing backdoor attack, but fail to defend the other two more solid attacks\footnote{$ASR \ge 40\%$ has been considered to be a discouraged level of defense performance against backdoor attacks~\cite{IMC}.}, TroJanNN and IMC, whose triggers are combined with artificially crafted noises and harder to be erased.
Especially, NAD achieves competitive performance to \sysname while processing some simple learning tasks, e.g., MNIST and Fashion-MNIST. 
However, when more complicated learning tasks, i.e., CIFAR-10 and CIFAR-100, are chosen, NAD only lowers about 40\%-60\% ASR of the IMC attack, which means NAD is still not robust enough as being faced with the state-of-the-art backdoor attack.
In sharp contrast, the \sysname successfully decrease the ASR from almost 100\% to be only about 10\% in all four datasets, which verifies the effectiveness of machine unlearning to be applied in backdoor defense.

Besides, to further validate the robustness of the defense methods against attacks, we implement diverse attack methods with varied trigger sizes.
Interestingly, as the trigger size increases, \sysname takes a different potential change trend on defense performance compared with other baselines.
For \sysname, its performance potentially becomes better along with the increasing of trigger sizes, however, other methods tends to perform worse in the same condition.
Diving to the bottom, the phenomenon is caused by the fact that larger trigger sizes can form more distinguished features and more deeply affect more neurons. 
For Fine-Pruning, the fact means more neurons required to be cleansed. 
For fine-tuning and NAD, deeper memories of neurons make polluted neurons harder to be purified.
Alternatively, a larger size of triggers weakens the performance of the existing backdoor defense methods.
Conversely, machine unlearning is sensitive to the feature distribution required to be unlearned.
According to Eq.~\ref{eq_unlearning}, more distinguished trigger patterns will promote the gradient ascent of machine unlearning to converge faster towards the desired direction and mitigate its side-effect on other normal data (less punishment).
Thus, \sysname tends to achieve better defense effect with increased trigger sizes. 
The characteristic further validates that \sysname is more robust than the previous methods in a wide range of applications.

\subsection{Further Understanding of \sysname}
Next, we further detail the tunable parameters that influence the performance of \sysname and discuss about the effectiveness of \sysname with extensive experiments.

\begin{figure*}[!ht]
	\centering
	\subfigure[MNIST]{
	\includegraphics[width=0.224\textwidth]{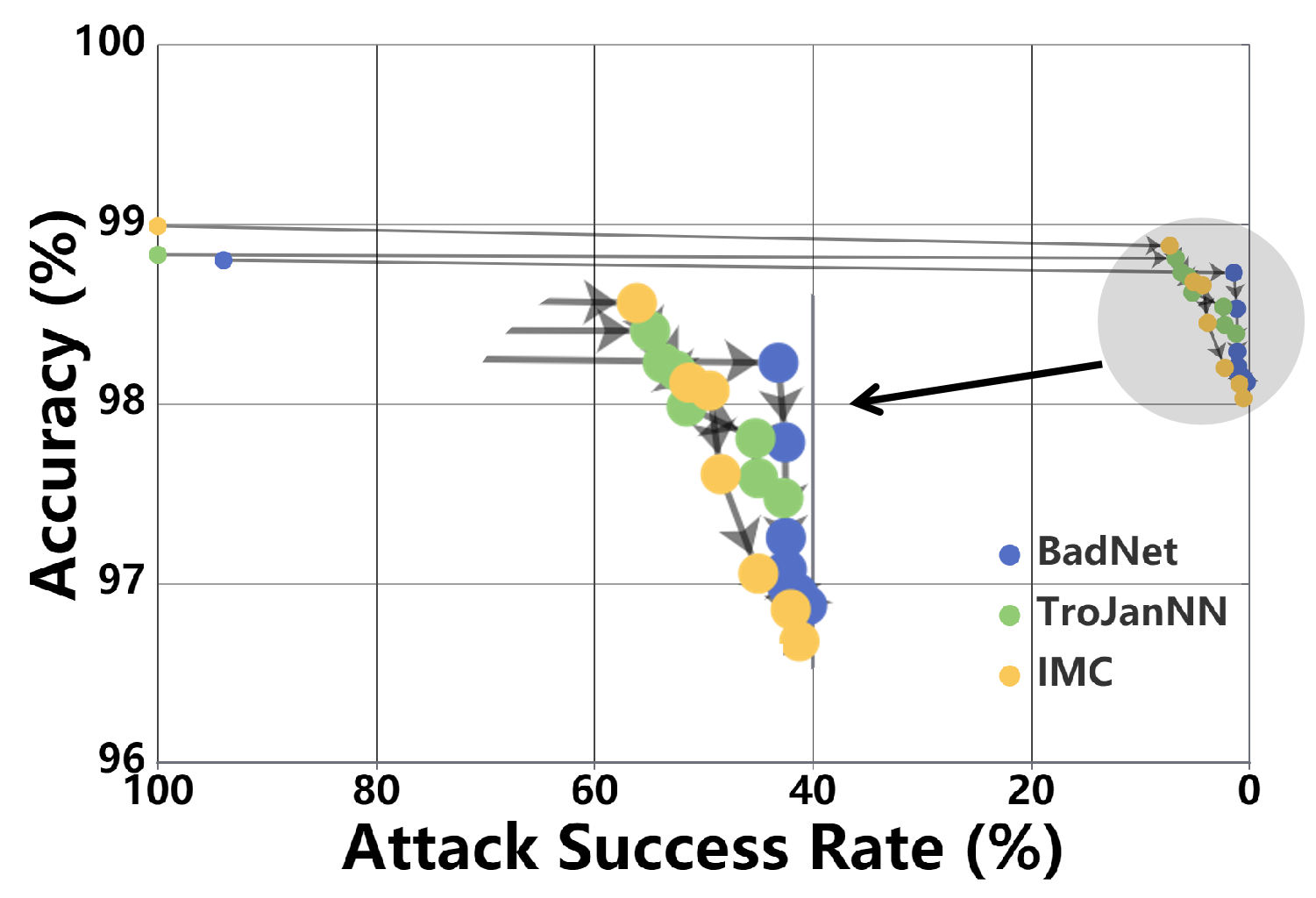} }
	\subfigure[Fashion-MNIST]{ \includegraphics[width=0.224\textwidth]{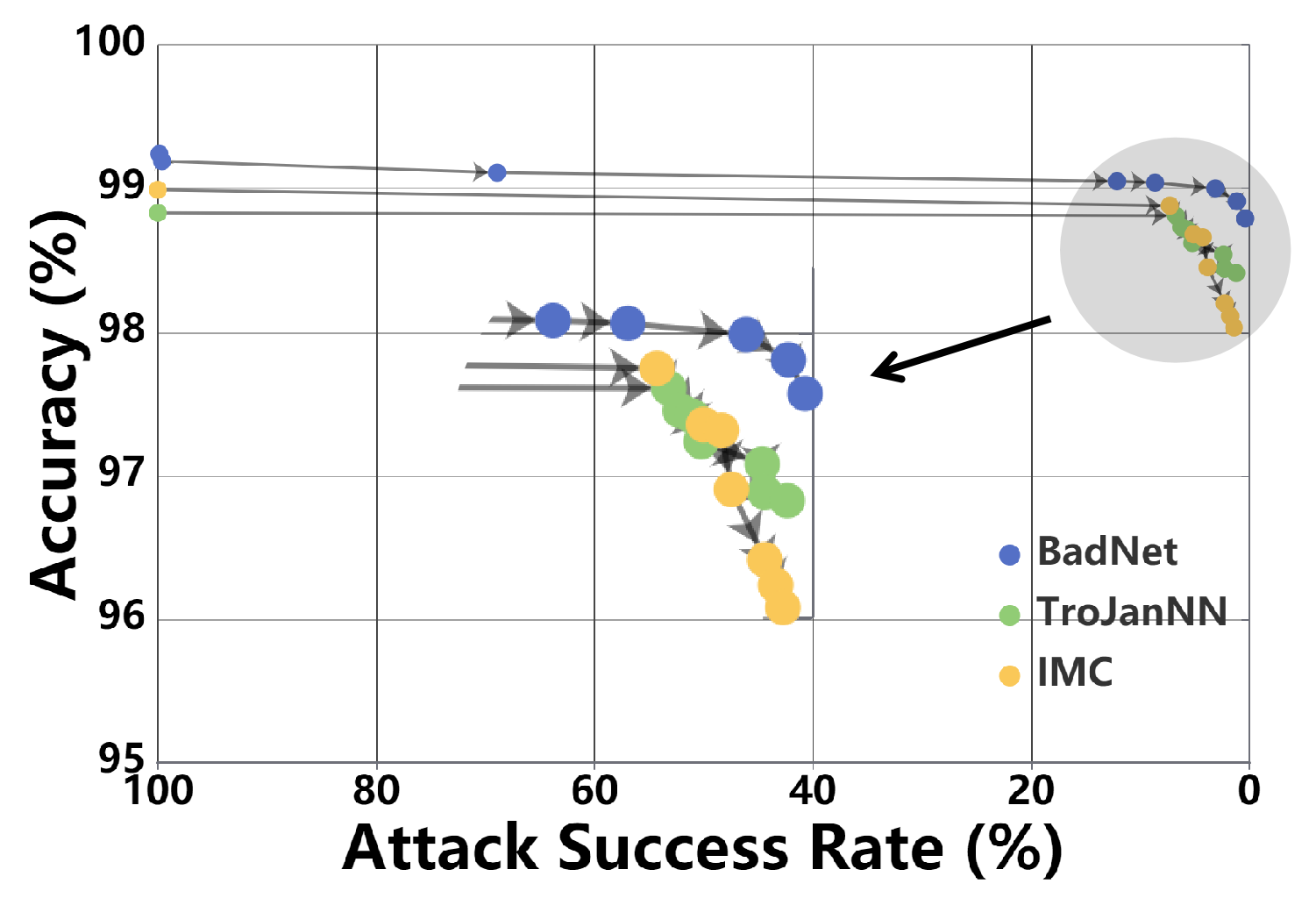} }
	\subfigure[CIFAR-10]{ \includegraphics[width=0.224\textwidth]{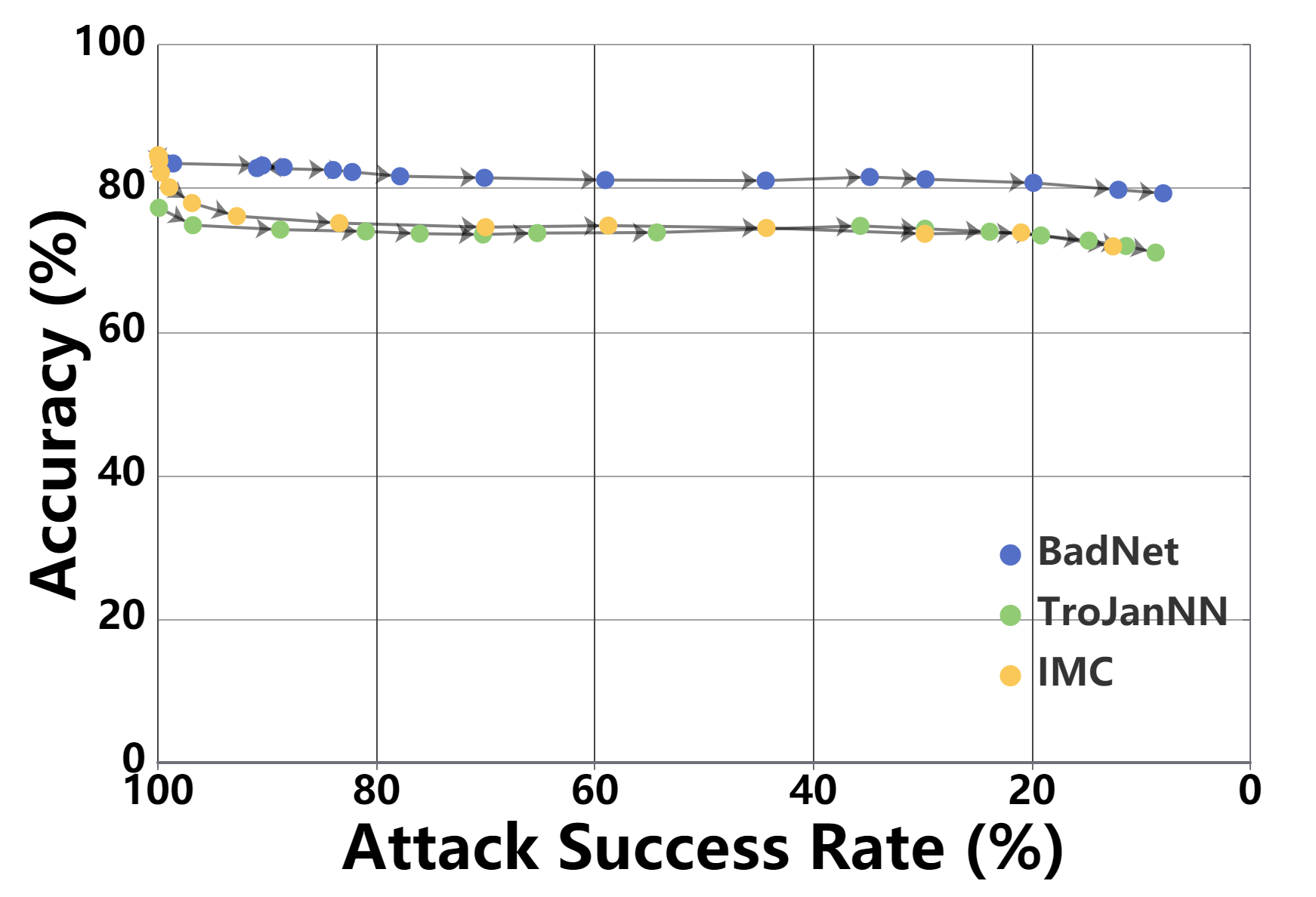} }
	\subfigure[CIFAR-100]{ \includegraphics[width=0.224\textwidth]{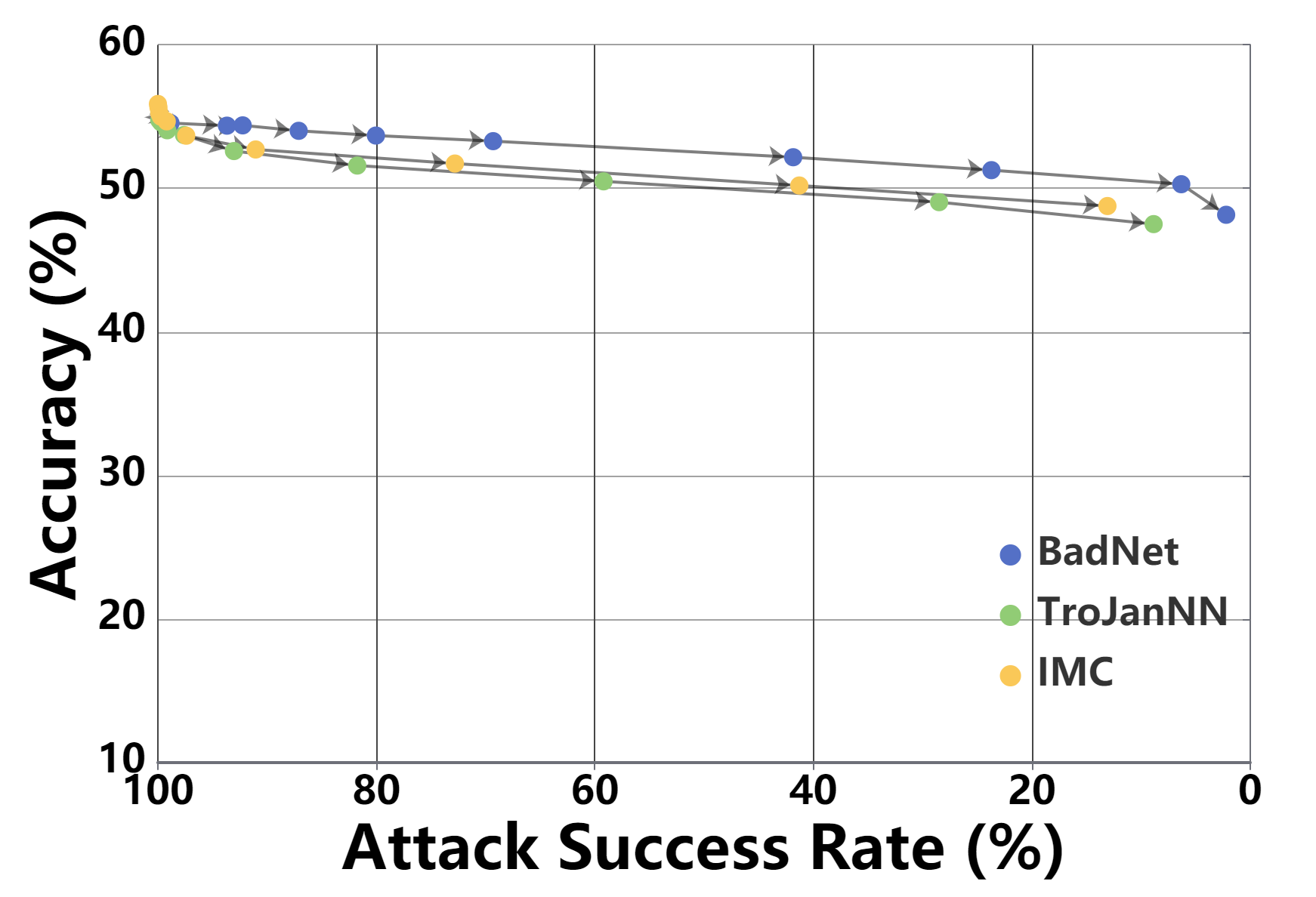} }
	\caption{Detailed iteration of our method against BadNet, TroJanNN, and IMC with MNIST, Fashion-MNIST, CIFAR-10, and CIFAR-100.
    The lower the slope of the lines in the above figures is, the better the performance of our method is.}
	\label{fig_detailed_iteration}
\end{figure*}

\subsubsection{Trade-off between ASR and Acc}
As the two most prominent metrics to evaluate a backdoor defense method, the correlation between Acc and ASR should be first concerned to understand our method. 
Fig.~\ref{fig_detailed_iteration} demonstrates the changing trend of Acc along with the increased ASR as the unlearning process is progressed (the flatter the curve, the better the defense effect).
It can be observed that there exists a trade-off between ASR and Acc, i.e., lower ASR requires more Acc sacrifice, and the sacrificed Acc intensively grows with the decreasing of ASR.
In other words, the reduction of ASR at the later iteration entails more drops of Acc than the earlier iteration.
This is because more unlearning iterations will deepen the side-effect of catastrophic forgetting~\cite{liu2020learn}, which further leads to the exploding gradient problem and expedite the Acc decline of the target model.
Besides, although Acc decay rate increases with unlearning iterations, we can still maintain less than 10\% Acc drops as the ASR is decreased to be almost 0\%.
Therefore, the phenomenon mentioned above will not weaken the practicality of \sysname in applications.



\begin{table*} [!h]
 \footnotesize
\centering 
\subtable[CIFAR-10]{  
\begin{tabular}{@{}c|c|cc|cc|cc|cc|cc@{}}
\toprule
\multirow{2}{*}{\textbf{Backdoor Attack}} & \multirow{2}{*}{\textbf{Trigger Size}} & \multicolumn{2}{c|}{\textbf{Before}} & \multicolumn{2}{c|}{\textbf{Fine-Pruning~\cite{fine_pruning}}} & \multicolumn{2}{c|}{\textbf{NAD~\cite{NAD}}} & \multicolumn{2}{c|}{\textbf{Fine-Tuning~\cite{qiao2019defending}}} & \multicolumn{2}{c}{\textbf{\sysname}} \\ \cmidrule(l){3-12} 
 &  & \textbf{ASR} & \textbf{Acc} & \textbf{ASR} & \textbf{Acc} & \textbf{ASR} & \textbf{Acc} & \textbf{ASR} & \textbf{Acc} & \textbf{ASR} & \textbf{Acc} \\ \midrule
\multirow{3}{*}{\textbf{BadNet \cite{Badnet}}} & \textbf{3$\times$3} & 98.80 & 83.81 & 32.07 & 77.96 & \textbf{4.24} & 69.00 & 9.52 & 78.95 & 7.96 & \textbf{79.31} \\
 & \textbf{5$\times$5} & 97.99 & 83.64 & 37.41 & 78.34 & 7.68 & 69.25 & 9.32 & 78.51 & \textbf{5.64} & \textbf{79.59} \\
 & \textbf{7$\times$7} & 98.01 & 82.89 & 35.24 & 76.29 & 8.95 & 70.49 & 10.77 & 78.32 & \textbf{4.66} & \textbf{79.19} \\ \midrule
\multirow{3}{*}{\textbf{TroJanNN \cite{Trojannn}}} & \textbf{3$\times$3} & 99.74 & 84.15 & 60.78 & 66.99 & 29.39 & 68.95 & 58.21 & 65.73 & \textbf{8.65} & \textbf{71.05} \\
 & \textbf{5$\times$5} & 99.52 & 83.84 & 56.89 & 66.55 & 38.76 & 66.80 & 57.80 & 66.32 & \textbf{6.15} & \textbf{74.12} \\
 & \textbf{7$\times$7} & 99.76 & 83.87 & 57.48 & 66.39 & 24.35 & 69.83 & 57.63 & 66.72 & \textbf{8.23} & \textbf{73.57} \\ \midrule
\multirow{3}{*}{\textbf{IMC \cite{IMC}}} & \textbf{3$\times$3} & 99.98 & 84.30 & 65.87 & 69.34 & 42.77 & 70.04 & 63.67 & 68.31 & \textbf{12.55} & \textbf{71.92} \\
 & \textbf{5$\times$5} & 99.70 & 84.62 & 69.13 & 63.68 & 48.18 & 70.89 & 67.64 & 69.77 & \textbf{11.01} & \textbf{71.11} \\
 & \textbf{7$\times$7} & 99.82 & 84.19 & 72.46 & 70.75 & 51.22 & 70.16 & 74.85 & 67.00 & \textbf{8.81} & \textbf{72.52} \\ \midrule
\multicolumn{2}{c|}{\textbf{Average}} & 99.26 & 83.92 & 54.15 & 70.70 & 29.17 & 69.49 & 45.49 & 71.07 & \textbf{8.18} & \textbf{74.71} \\ \midrule
\end{tabular}
       \label{tab:thirdtable}  
}
\qquad
  \subtable[CIFAR-100]{          
\begin{tabular}{@{}c|c|cc|cc|cc|cc|cc@{}}
\toprule
\multirow{2}{*}{\textbf{Backdoor Attack}} & \multirow{2}{*}{\textbf{Trigger Size}} & \multicolumn{2}{c|}{\textbf{Before}} & \multicolumn{2}{c|}{\textbf{Fine-Pruning~\cite{fine_pruning}}} & \multicolumn{2}{c|}{\textbf{NAD~\cite{NAD}}} & \multicolumn{2}{c|}{\textbf{Fine-Tuning~\cite{qiao2019defending}}} & \multicolumn{2}{c}{\textbf{\sysname}} \\ \cmidrule(l){3-12} 
 &  & \textbf{ASR} & \textbf{Acc} & \textbf{ASR} & \textbf{Acc} & \textbf{ASR} & \textbf{Acc} & \textbf{ASR} & \textbf{Acc} & \textbf{ASR} & \textbf{Acc} \\ \midrule
\multirow{3}{*}{\textbf{BadNet \cite{Badnet}}} & \textbf{3$\times$3} & 98.06 & 54.40 & 47.58 & 48.12 & 3.09 & 33.33 & 20.19 & 47.37 & \textbf{2.20} & \textbf{48.16} \\
 & \textbf{5$\times$5} & 98.44 & 54.79 & 51.28 & 44.64 & 4.56 & 31.89 & 13.75 & 46.98 & \textbf{3.65} & \textbf{47.81} \\
 & \textbf{7$\times$7} & 97.94 & 55.17 & 46.31 & 43.57 & 4.73 & 34.27 & 17.71 & 46.11 & \textbf{3.81} & \textbf{49.11} \\ \midrule
\multirow{3}{*}{\textbf{TroJanNN \cite{Trojannn}}} & \textbf{3$\times$3} & 99.85 & 54.23 & 71.29 & 45.17 & 26.35 & 32.24 & 61.23 & 43.51 & \textbf{8.84} & \textbf{47.51} \\
 & \textbf{5$\times$5} & 99.49 & 54.15 & 60.45 & 43.44 & 19.63 & 32.80 & 60.73 & 45.28 & \textbf{8.76} & \textbf{48.13} \\
 & \textbf{7$\times$7} & 99.15 & 53.50 & 56.62 & 45.41 & 17.05 & 31.08 & 59.40 & 43.86 & \textbf{5.20} & \textbf{48.59} \\ \midrule
\multirow{3}{*}{\textbf{IMC \cite{IMC}}} & \textbf{3$\times$3} & 99.99 & 55.50 & 86.51 & 46.29 & 71.83 & 32.43 & 77.65 & 42.66 & \textbf{13.07} & \textbf{48.77} \\
 & \textbf{5$\times$5} & 99.88 & 55.54 & 87.52 & 42.87 & 54.98 & 31.64 & 68.41 & 47.46 & \textbf{13.53} & \textbf{48.36} \\
 & \textbf{7$\times$7} & 99.72 & 55.76 & 83.17 & 45.36 & 63.66 & 31.97 & 64.82 & 44.84 & \textbf{10.05} & \textbf{48.07} \\ \midrule
\multicolumn{2}{c|}{\textbf{Average}} & 99.17 & 54.78 & 65.64 & 44.98 & 32.88 & 32.41 & 49.32 & 45.34 & \textbf{7.68} & \textbf{48.28} \\ \midrule
\end{tabular}
       \label{tab:fourthtable}  
}
\caption{Comparison analysis with CIFAR-10, and CIFAR-100.}
\label{table_comparison_with_cifar}
\end{table*}

\subsubsection{Impact of Holding Ratio}
\begin{figure*}[!h]
	\centering
	\subfigure[Acc on MNIST]{ \includegraphics[width=0.22\textwidth]{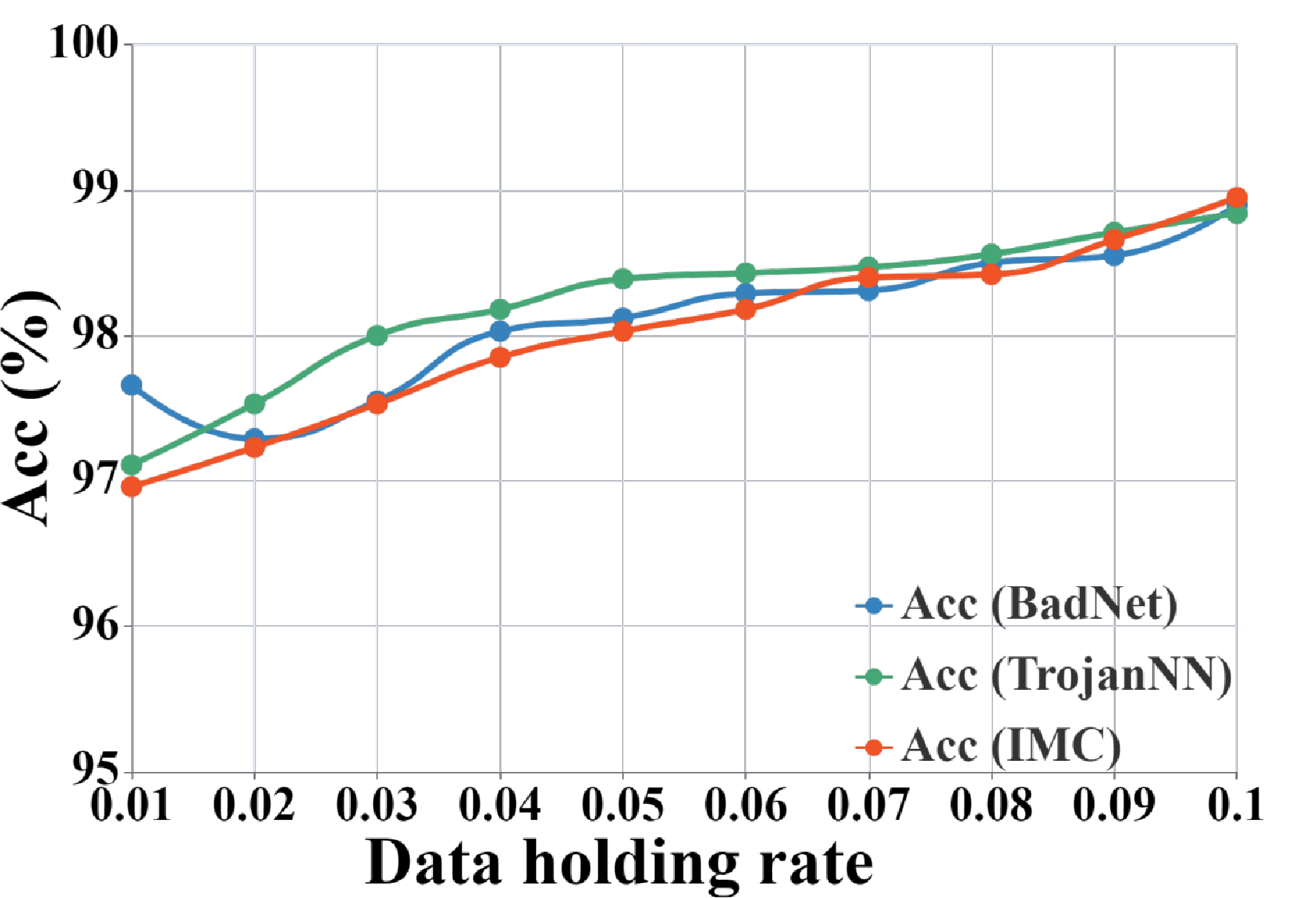} }
	\subfigure[Acc on Fashion-MNIST]{ \includegraphics[width=0.22\textwidth]{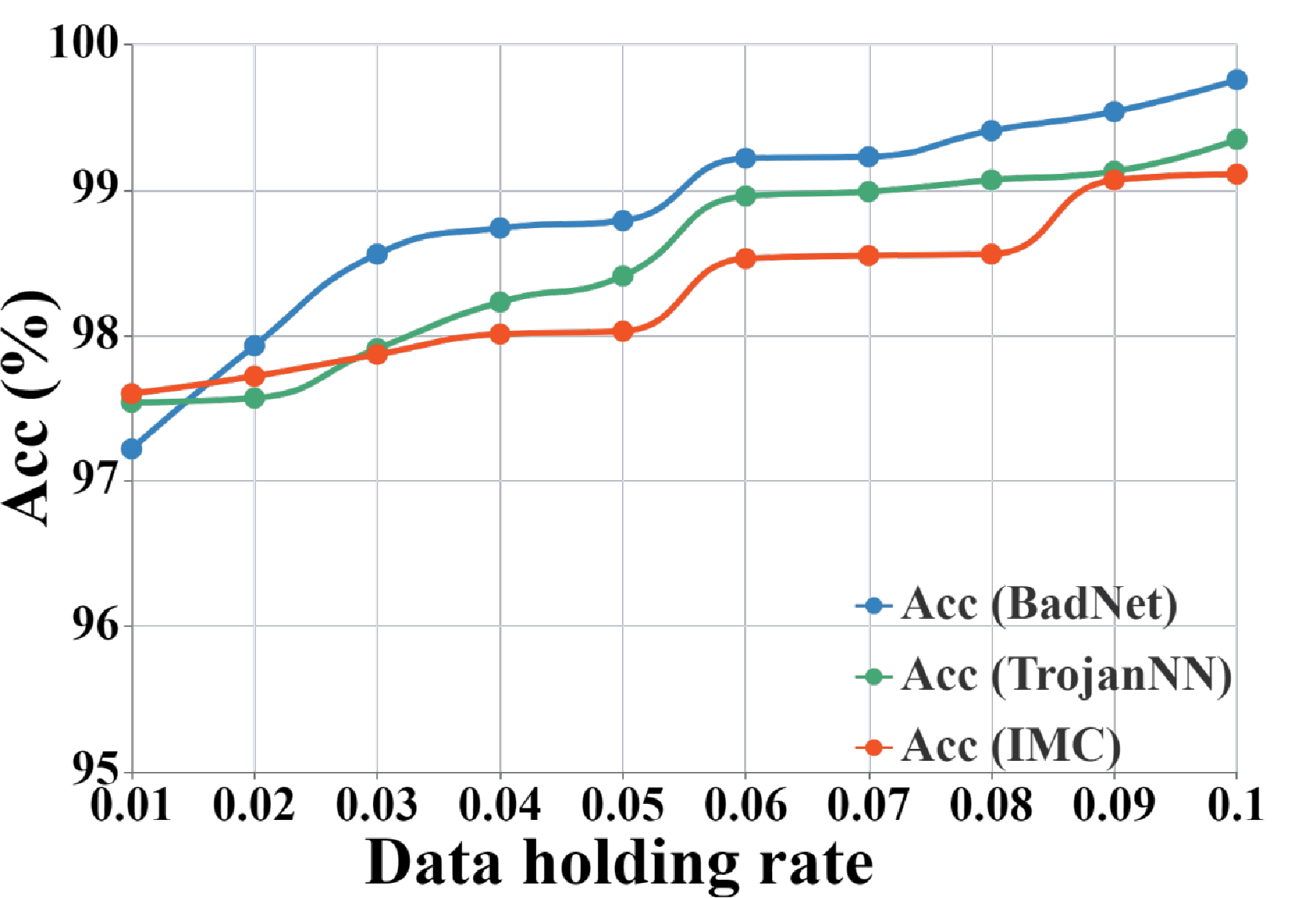} }
	\subfigure[Acc on CIFAR-10]{ \includegraphics[width=0.22\textwidth]{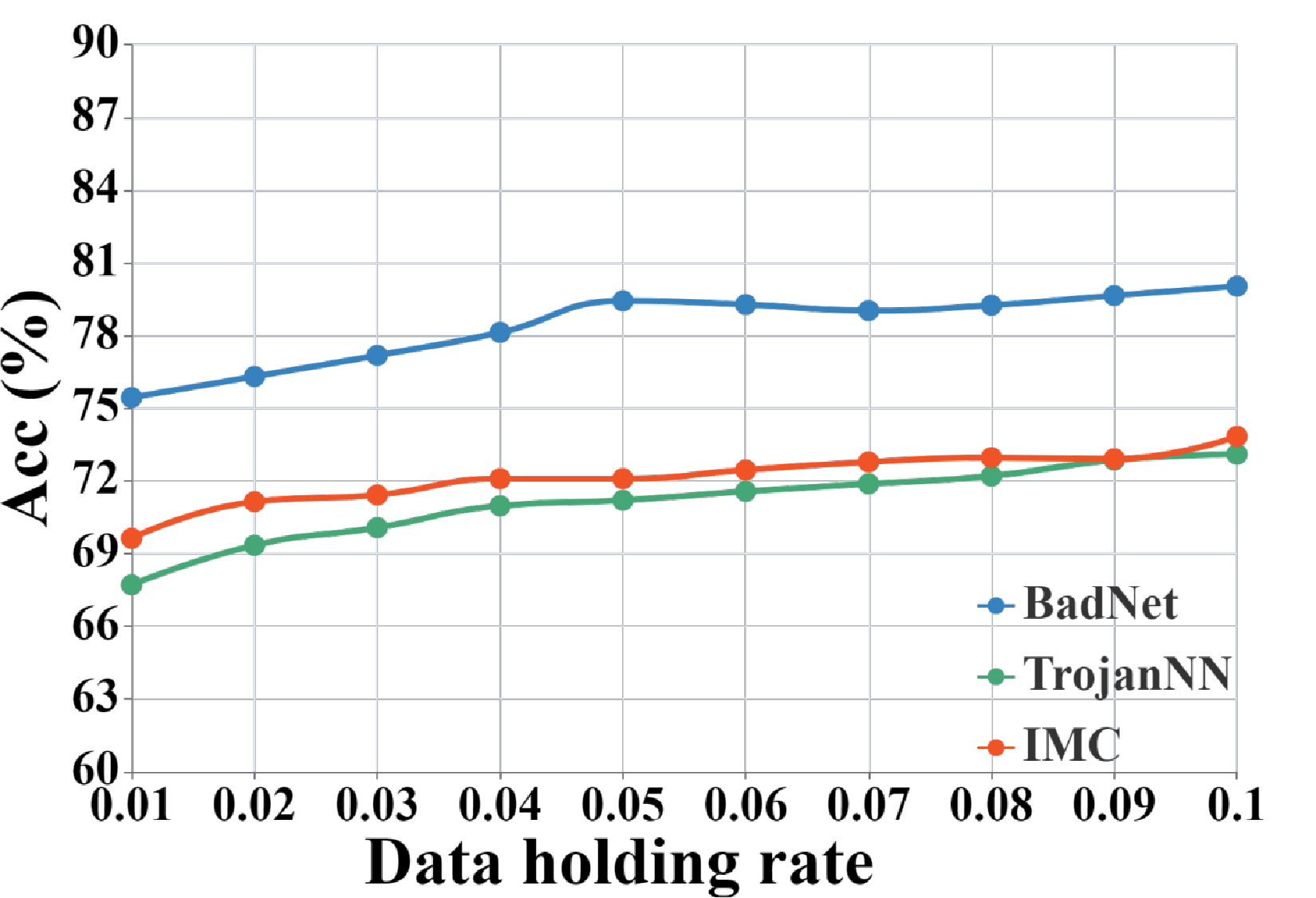} }
	\subfigure[Acc on CIFAR-100]{ \includegraphics[width=0.22\textwidth]{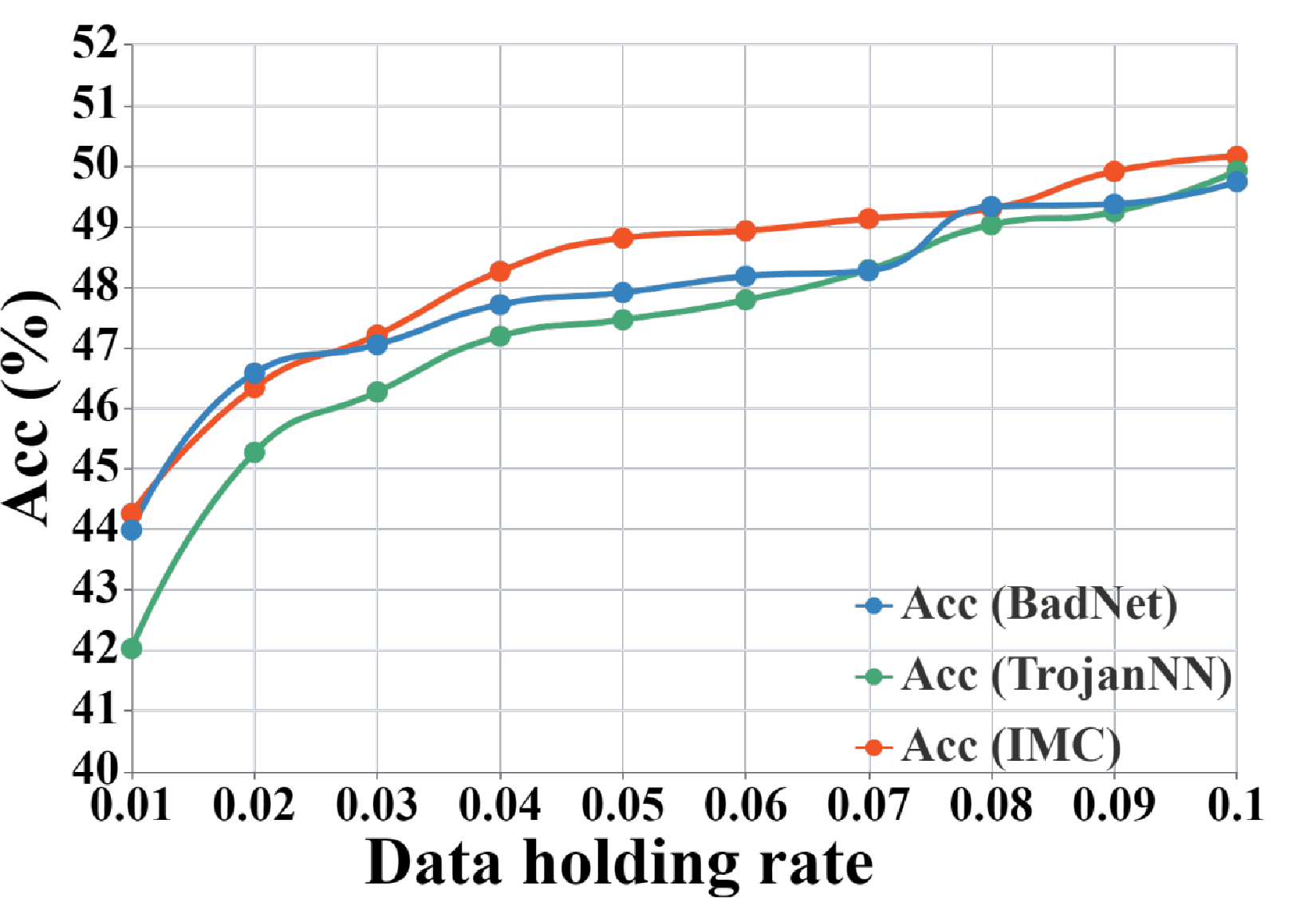} }
	\subfigure[ASR on MNIST]{ \includegraphics[width=0.22\textwidth]{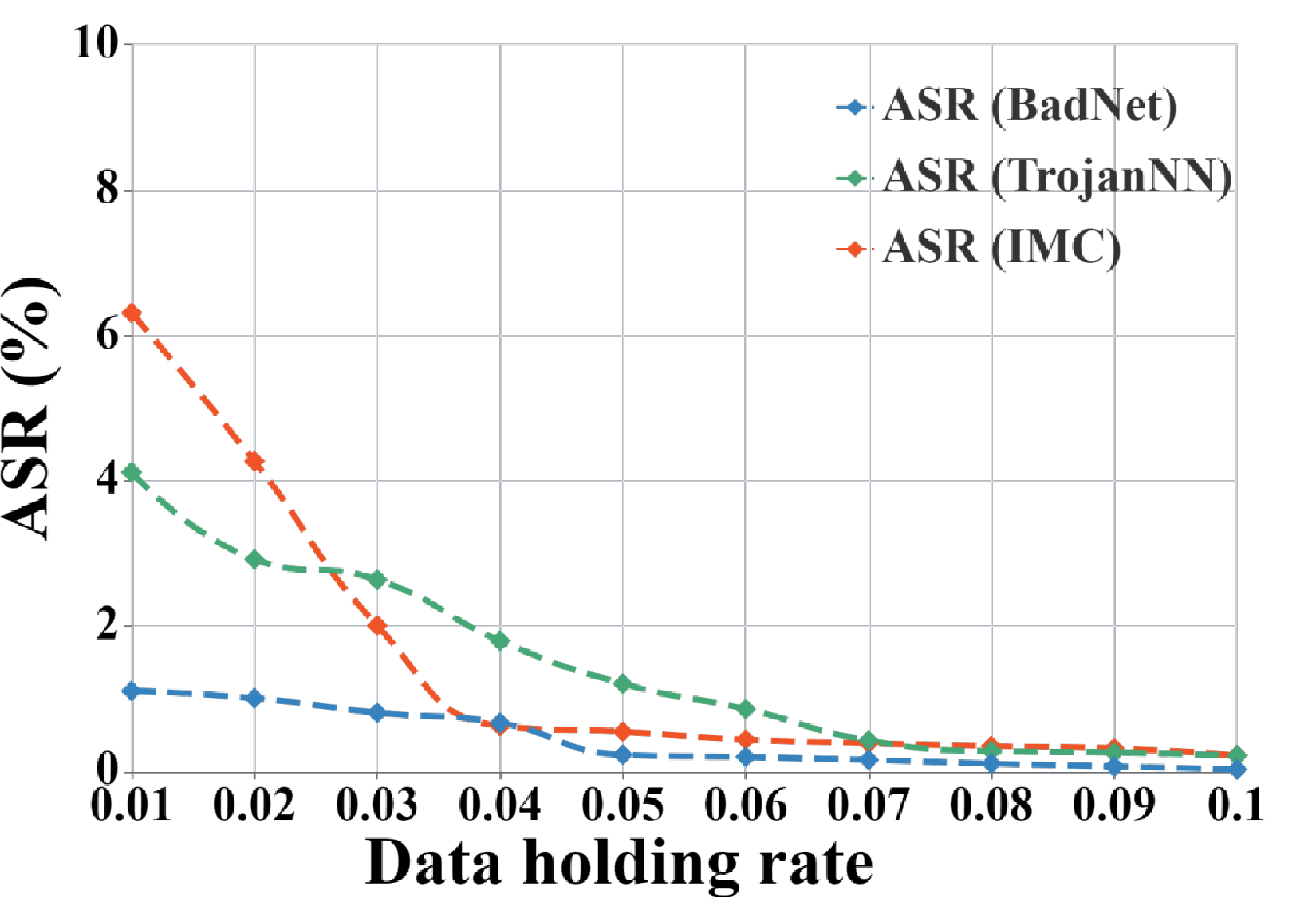} }
	\subfigure[ASR on Fashion-MNIST]{ \includegraphics[width=0.22\textwidth]{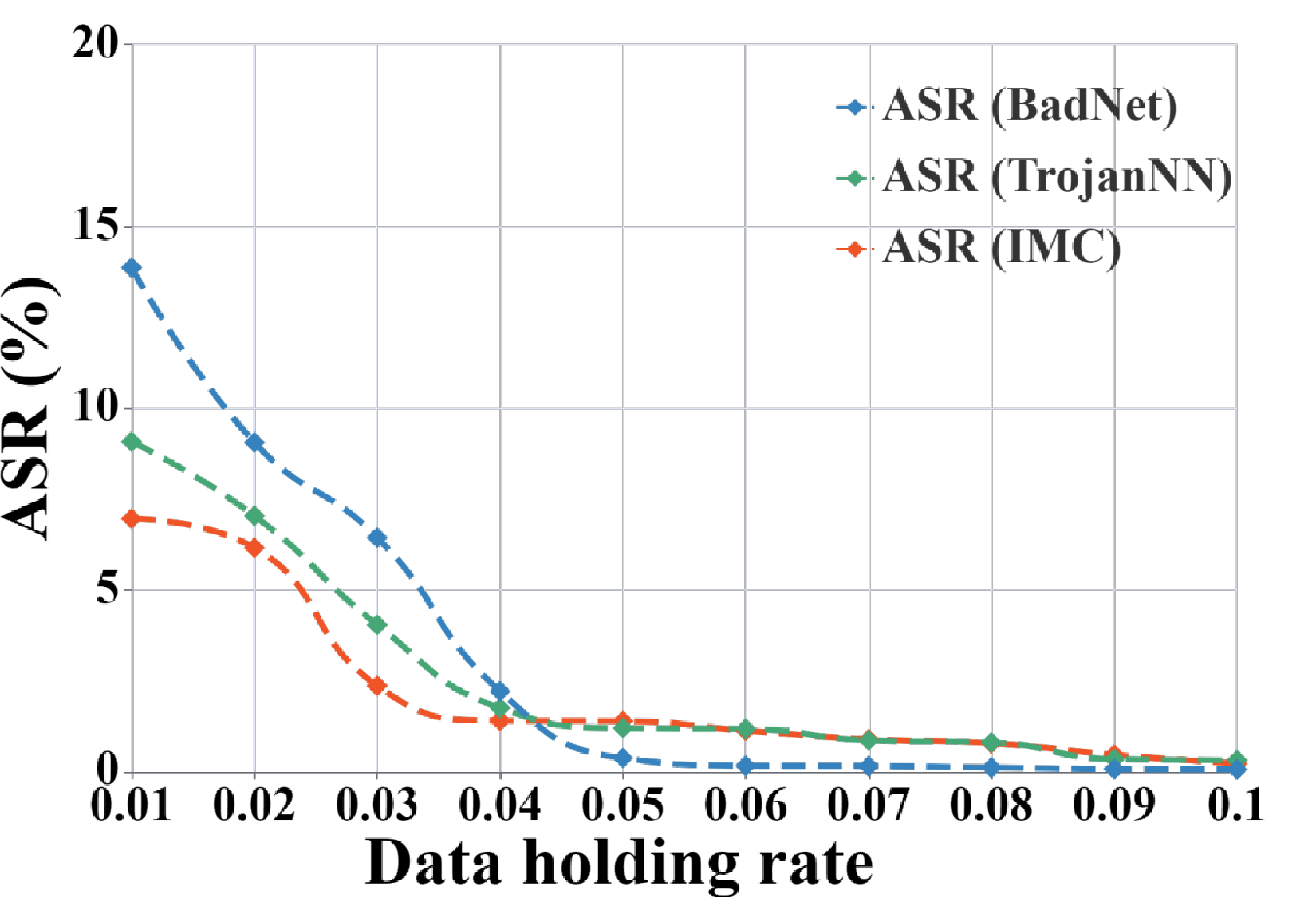} }
	\subfigure[ASR on CIFAR-10]{ \includegraphics[width=0.22\textwidth]{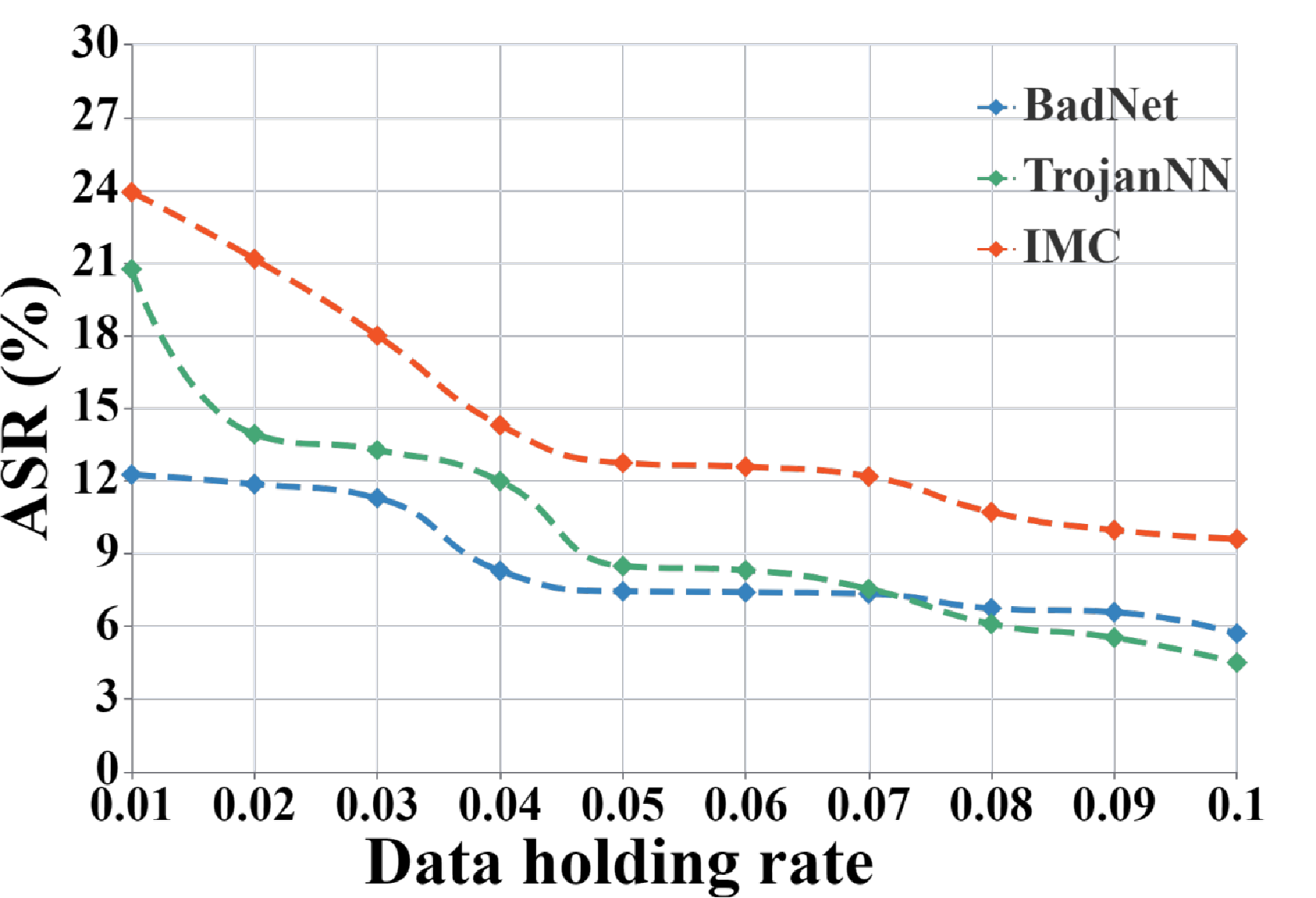} }
	\subfigure[ASR on CIFAR-100]{ \includegraphics[width=0.22\textwidth]{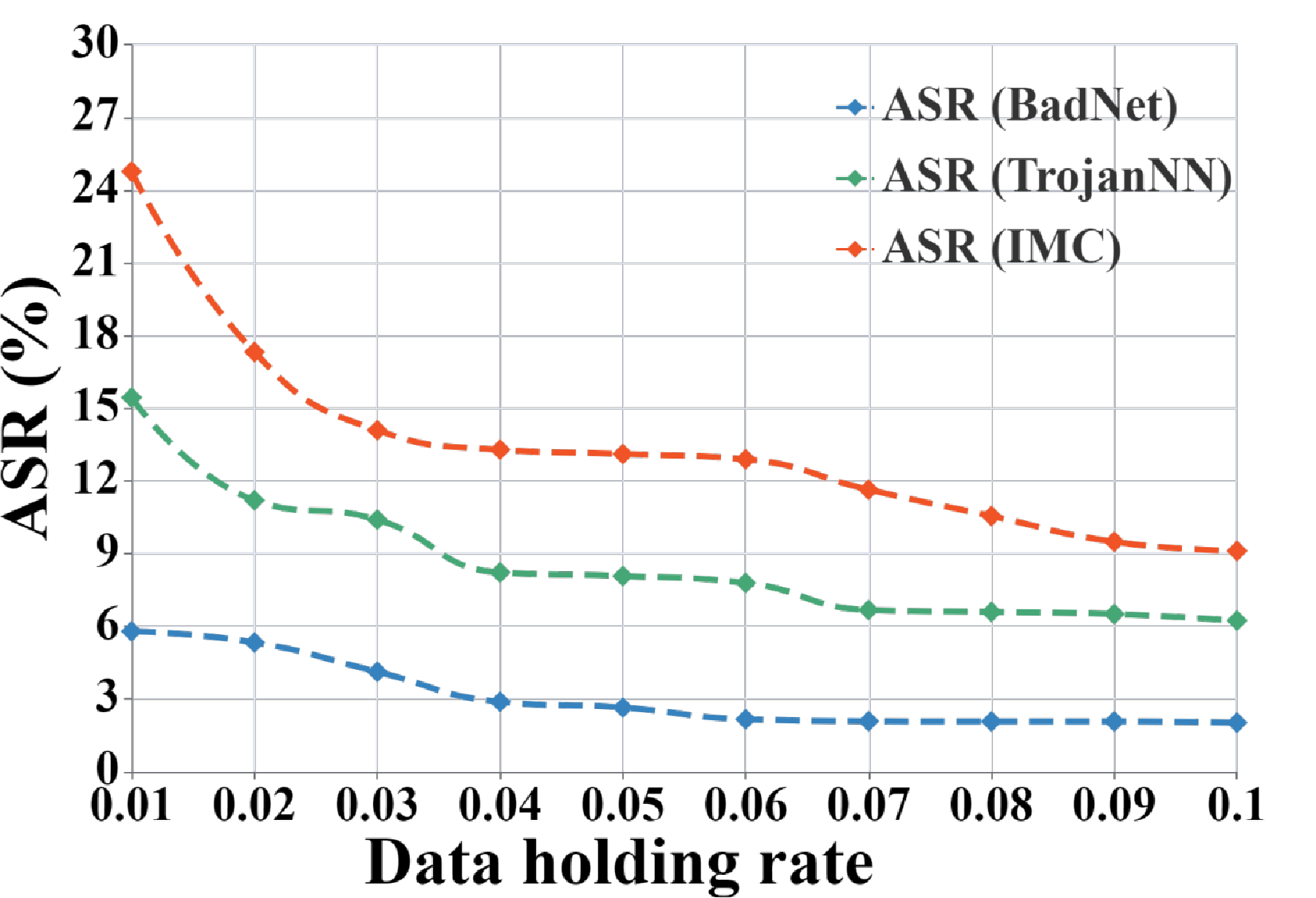} }
	\caption{The performance (measured by Acc and ASR) of our method with different ratios of the clean data set.}
	\label{fig_data_ratio}
\end{figure*}

As discussed in Section~\ref{sec_problem}, another critical factor that affects the practicality of \sysname in applications is the number of clean data held by the defender.
Here, the clean data specify the testing (validation data) that are not used in model training, and the holding ratio~\cite{NAD} is computed by: $\frac{n_{clean}}{n_{training}}\times 100\%$.
Ideally, if the defender holds the enough clean data, he can simply retrain a clean model or forcibly fine-tune the backdoored model.
Unfortunately, the harsh condition can be hardly met in most real-world applications as mentioned in the prior work~\cite{fine_pruning, qiao2019defending, NAD}.
Here, to further validate the robustness of \sysname, Fig.~\ref{fig_data_ratio} illustrated the defense performance change of \sysname under different ratios (from 1\% to 10\%) of clean data held by the defender.
Intuitively, the experimental results are within expectation.
\sysname performs better with higher holding ratios of clean data.
Furthermore, even with only 1\% clean data (500 clean data points), \sysname still achieves an appealing defense rate (lowering at most 98\% ASR).


\begin{table*} [!h]
\centering 
\footnotesize
  \subtable[MNIST]{          
\begin{tabular}{@{}c|cc|cc|cc@{}}
\toprule
\multirow{2}{*}{\textbf{$\beta/\alpha$}} &
  \multicolumn{2}{c|}{\textbf{BadNet}} &
  \multicolumn{2}{c|}{\textbf{TrojanNN}} &
  \multicolumn{2}{c}{\textbf{IMC}} \\ \cmidrule(l){2-7} 
 &
  \textbf{Accuracy} &
  \textbf{ASR} &
  \textbf{Accuracy} &
  \textbf{ASR} &
  \textbf{Accuracy} &
  \textbf{ASR} \\ \midrule
\textbf{0.001} & 93.87 ($\downarrow$5) & \textbf{0 ($\downarrow$99.52)} & 91.77 ($\downarrow$7.05) & \textbf{0 ($\downarrow$100)} & 89.51 ($\downarrow$9.49) & 0.11 ($\downarrow$99.89) \\
\textbf{0.01} & 95.25 ($\downarrow$3.62) & \textbf{0 ($\downarrow$99.52)} & 93.59 ($\downarrow$5.23) & \textbf{0 ($\downarrow$100)} & 92.13 ($\downarrow$6.87) & 0.22 ($\downarrow$99.78) \\
\textbf{0.1} & 96.27 ($\downarrow$2.60) & \textbf{0 ($\downarrow$99.52)} & 95.75 ($\downarrow$3.07) & 1.11 ($\downarrow$98.89) & 94.12 ($\downarrow$4.88) & 0.41 ($\downarrow$99.59) \\
\textbf{1} & 98.12 ($\downarrow$0.75) & 0.23 ($\downarrow$99.29) & 98.39 ($\downarrow$0.43) & 1.21 ($\downarrow$98.79) & 98.03 ($\downarrow$0.97) & 0.55 ($\downarrow$99.45) \\
\textbf{10} & 98.77 ($\downarrow$0.10) & 0.44 ($\downarrow$99.08) & 98.41 ($\downarrow$0.41) & 1.56 ($\downarrow$98.44) & 98.31 ($\downarrow$0.69) & 0.72 ($\downarrow$99.28) \\
\textbf{100} & \textbf{98.82 ($\downarrow$0.05)} & 0.51 ($\downarrow$99.01) & \textbf{98.46 ($\downarrow$0.36)} & 1.77 ($\downarrow$98.23) & \textbf{98.49 ($\downarrow$0.51)} & 0.78 ($\downarrow$99.22) \\ \bottomrule
\end{tabular}
      \label{tab:secondtable}  
}
\subtable[CIFAR-10]{  
\begin{tabular}{@{}c|cc|cc|cc@{}}
\toprule
\multirow{2}{*}{\textbf{$\beta/\alpha$}} &
  \multicolumn{2}{c|}{\textbf{BadNet}} &
  \multicolumn{2}{c|}{\textbf{TrojanNN}} &
  \multicolumn{2}{c}{\textbf{IMC}} \\ \cmidrule(l){2-7} 
 &
  \textbf{Accuracy} &
  \textbf{ASR} &
  \textbf{Accuracy} &
  \textbf{ASR} &
  \textbf{Accuracy} &
  \textbf{ASR} \\ \midrule
\textbf{0.001} &
  77.76 ($\downarrow$6.05) &
  2.2 ($\downarrow$96.6) &
  62.81 ($\downarrow$21.34) &
  4.36 ($\downarrow$94.44) &
  64.56 ($\downarrow$19.74) &
  1.03 ($\downarrow$98.95) \\
\textbf{0.01} &
  78.37 ($\downarrow$5.44) &
  3.91 ($\downarrow$94.89) &
  66.11 ($\downarrow$18.04) &
  7.75 ($\downarrow$91.05) &
  66.53 ($\downarrow$17.77) &
  7.35 ($\downarrow$92.63) \\
\textbf{0.1} &
  79.17 ($\downarrow$4.64) &
  4.69 ($\downarrow$94.11) &
  68.47 ($\downarrow$15.68) &
  8.43 ($\downarrow$90.37) &
  68.13 ($\downarrow$16.17) &
  9.29 ($\downarrow$90.69) \\
\textbf{1} &
  79.09 ($\downarrow$4.72) &
  7.88 ($\downarrow$90.92) &
  71.25 ($\downarrow$12.9) &
  9.45 ($\downarrow$89.35) &
  71.89 ($\downarrow$12.41) &
  11.52 ($\downarrow$88.46) \\
\textbf{10} &
  79.57 ($\downarrow$4.24) &
  9.36 ($\downarrow$89.44) &
  73.01 ($\downarrow$11.14) &
  27.69 ($\downarrow$71.11) &
  72.32 ($\downarrow$11.98) &
  15.68 ($\downarrow$84.3) \\
\textbf{100} &
  \textbf{80.47 ($\downarrow$3.34)} &
  9.99 ($\downarrow$88.81) &
  \textbf{75.63 ($\downarrow$8.52)} &
  30.39 ($\downarrow$68.41) &
  \textbf{73.17 ($\downarrow$11.13)} &
  20.17 ($\downarrow$79.81) \\ \bottomrule
\end{tabular}
      \label{tab:firsttable}  
}
\qquad
  \subtable[CIFAR-100]{          
\begin{tabular}{@{}c|cc|cc|cc@{}}
\toprule
\multirow{2}{*}{\textbf{$\beta/\alpha$}} &
  \multicolumn{2}{c|}{\textbf{BadNet}} &
  \multicolumn{2}{c|}{\textbf{TrojanNN}} &
  \multicolumn{2}{c}{\textbf{IMC}} \\ \cmidrule(l){2-7} 
 &
  \textbf{Accuracy} &
  \textbf{ASR} &
  \textbf{Accuracy} &
  \textbf{ASR} &
  \textbf{Accuracy} &
  \textbf{ASR} \\ \midrule
\textbf{0.001} &
  44.69 ($\downarrow$9.71) &
  0.36 ($\downarrow$97.7) &
  38.87 ($\downarrow$15.36) &
  0.96 ($\downarrow$98.89) &
  45.5 ($\downarrow$10) &
  9.24 ($\downarrow$90.75) \\
\textbf{0.01} &
  46.37 ($\downarrow$8.03) &
  0.89 ($\downarrow$97.17) &
  41.35 ($\downarrow$12.88) &
  1.95 ($\downarrow$97.9) &
  46.11 ($\downarrow$9.39) &
  9.53 ($\downarrow$90.46) \\
\textbf{0.1} &
  47.73 ($\downarrow$6.67) &
  1.33 ($\downarrow$96.73) &
  46.63 ($\downarrow$7.60) &
  6.64 ($\downarrow$93.21) &
  48.27 ($\downarrow$7.23) &
  12.16 ($\downarrow$87.83) \\
\textbf{1} &
  48.59 ($\downarrow$5.81) &
  2.96 ($\downarrow$95.1) &
  48.09 ($\downarrow$6.14) &
  9.18 ($\downarrow$90.67) &
  49.01 ($\downarrow$6.49) &
  13.04 ($\downarrow$86.95) \\
\textbf{10} &
  50.55 ($\downarrow$3.85) &
  10.75 ($\downarrow$87.31) &
  48.69 ($\downarrow$5.54) &
  11.13 ($\downarrow$88.72) &
  50.63 ($\downarrow$4.87) &
  14.59 ($\downarrow$85.4) \\
\textbf{100} &
  \textbf{52.95 ($\downarrow$1.45)} &
  29.6 ($\downarrow$68.46) &
  \textbf{49.93 ($\downarrow$4.3)} &
  14.87 ($\downarrow$84.98) &
  \textbf{51.32 ($\downarrow$4.18)} &
  15.56 ($\downarrow$84.43) \\ \bottomrule
\end{tabular}
}
\caption{Performance change of \sysname with different ratios of $\alpha$ and $\beta$ with Fashion-MNIST and CIFAR-100.} 
\label{table_alpha_and_beta}
\end{table*}

\begin{figure*}[!ht]
	\centering
    \includegraphics[width=0.8\textwidth]{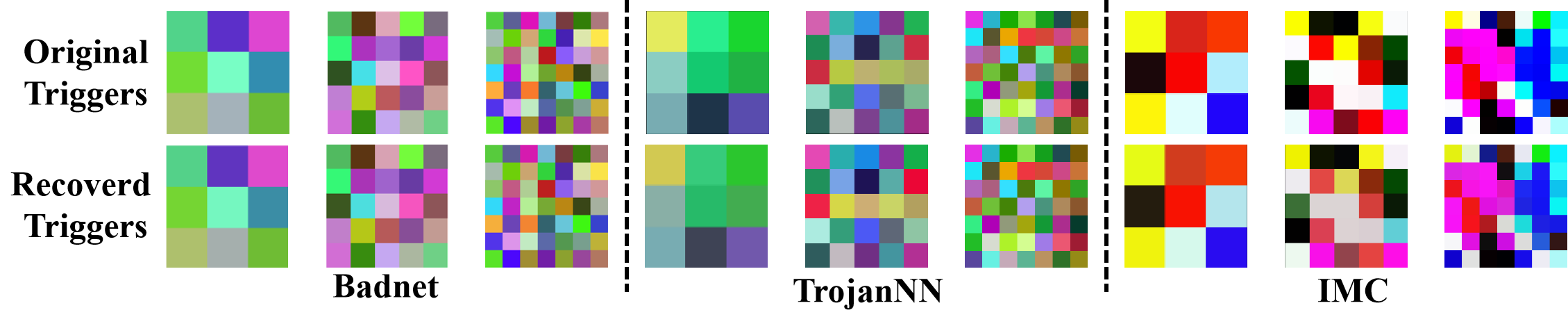}
	\caption{Comparison between the original trigger (top row) and the recovered triggers (bottom row). The triggers are recovered over BadNet, TrojanNN and IMC in CIFAR-10. The same trigger recovery operation is repeated for three times over three different triggers for generality.}
	\label{fig_recoverd_trigger}
\end{figure*}

\subsubsection{Impact of $\alpha$ and $\beta$}
According to Eq.~\ref{eq_unlearning}, two tunable coefficients $\alpha$ and $\beta$ in the unlearning loss $\mathcal{L}_{U}$ can control the behaviour of backdoor erasing for \sysname.
Specifically, the ratio of $\alpha$ to $\beta$ determines whether \sysname is inclined to maintain model performance or lower the ASR.
As ${\beta}/{\alpha}$ increases, more attention of \sysname is paid to control model performance loss, and otherwise, \sysname tends to lower the ASR as much as possible.
Such a conclusion is validated in the experiments shown in Table~\ref{table_alpha_and_beta} with Fashion-MNIST and CIFAR-100, which differs ${\beta}/{\alpha}$ from $0.0001$ to $100$ with fixed $\beta = 1$.
From the results, the performance change trend of \sysname accords with our anticipation.
On the one hand, when the ratio is chosen to be a very small value ($0.0001$), the penalty item is approximately canceled, so that ASR can be lowered to almost 0\% but more than 10\% Acc drop (catastrophic forgetting) is introduced.
On the other hand, the penalty item plays a major role and the Acc drop is well controlled when ${\beta}/{\alpha}$ reaches $100$.
Moreover, the above experiments also verify that the introduction of the dynamic penalty mechanism in Section~\ref{sec_approach} is valid to avoid catastrophic forgetting.



\subsubsection{Visualization of Recovered Triggers}
For our method, the similarity between the original trigger and the recovered trigger is the footstone to the effectiveness of unlearning.
In other words, if the recovered trigger patterns are not valid, the subsequent trigger unlearning becomes meaningless.
Here, for better understanding of why \sysname can work, we visualize the original trigger used by the attacker and the triggers synthesized by the defender in Fig~.\ref{fig_recoverd_trigger}.
From left to right, Fig.~\ref{fig_recoverd_trigger} plots the original trigger, and the triggers recovered by \sysname for three different attacks. 
Since the raw triggers with $3 \times 3$ resolution are not convenient for observation, we enlarge the image while keeping the aspect ratio.
Although the patterns of recovered triggers have a difference with the original trigger, the difference for most pixels is negligible.
As a result, the trigger pattern recovery method leveraged by \sysname can basically satisfy the requirement for backdoor defense but still spare space for improvements for a wider range of applications.


\begin{figure}[h!]
	\centering
	\includegraphics[width=0.4\textwidth]{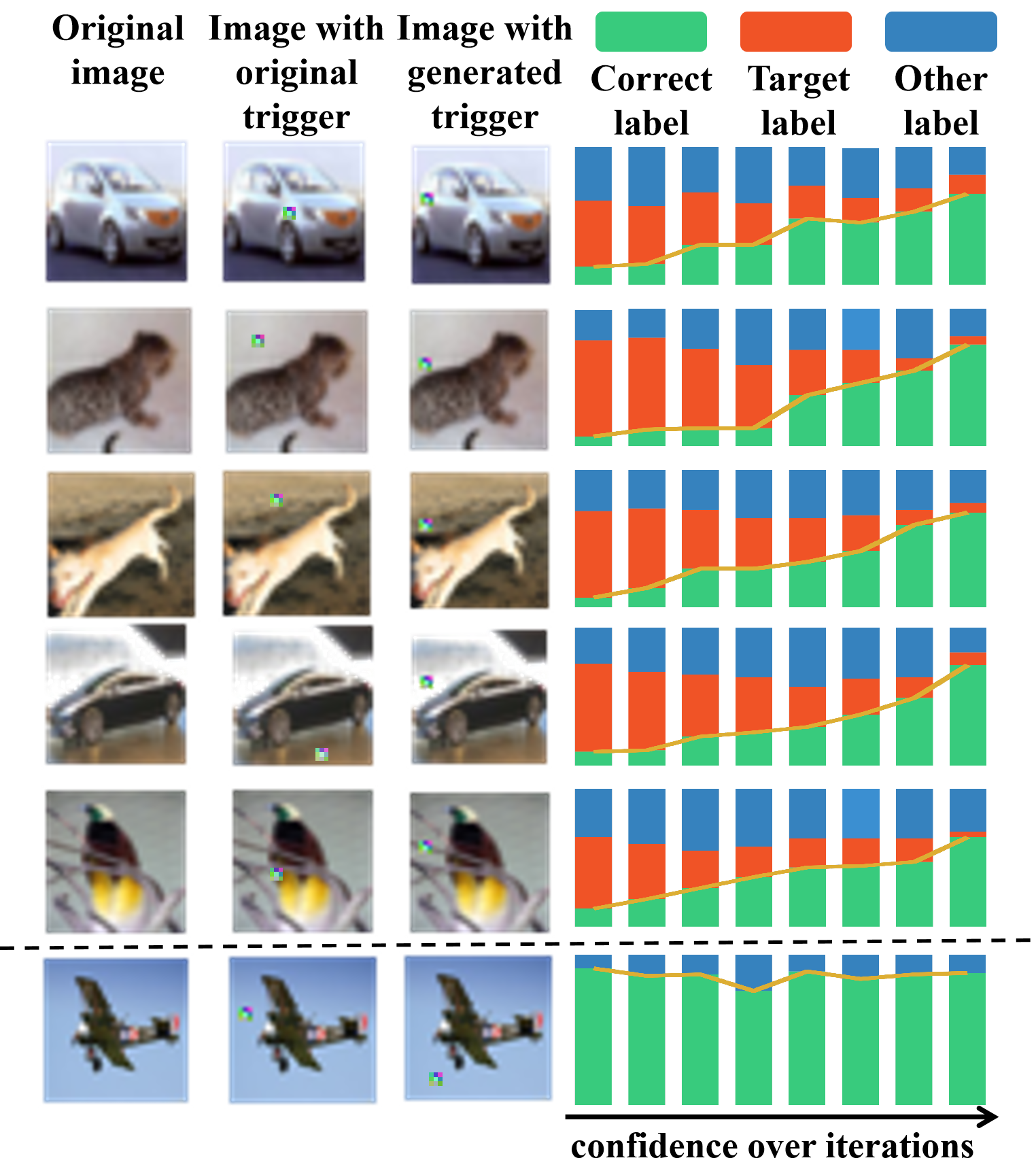}
	\caption{Examples of the detailed backdoor unlearning workflows over different images. The left part shows the original image, the images with triggers injected by the attacker and the images with triggers recovered by \sysname. The right part denotes the model prediction change about the triggered images in different unlearning iterations.}
	\label{fig_understanding_unlearning}
\end{figure}

\subsubsection{Understanding of Backdoor Unlearning}
To further understand the advantages of our backdoor unlearning algorithm, we plot the Fig.~\ref{fig_understanding_unlearning} to show the unlearning details of \sysname.
On the left of Fig.~\ref{fig_understanding_unlearning}, the three columns list clean images, images with the original trigger, and images with the recovered triggers, respectively.
On the right, the histograms demonstrate the prediction change of the victim model over the poisoned data during different iterations of the backdoor unlearning procedure.
Intuitively, we can derive the following insights about backdoor unlearning.

First, unlike NAD whose defense needs tens of iterations of distillation, backdoor unlearning is a more radical but effective defense method.
The model predictions on triggered inputs are steadily converged to the correct direction in only eight iterations.
Then, backdoor unlearning is robust to the change of trigger positions.
The intrinsic principle of NN makes it lose the perception of location information, which allows the attacker to launch attacks with randomly located triggers and results in the recovered triggers of \sysname being often attached to different places.
Nevertheless, no matter where the triggers are, backdoor unlearning can always succeed in erasing the polluted memories from the model.
Finally, backdoor unlearning hardly affects correct memories.
At the last row of Fig.~\ref{fig_understanding_unlearning}, we do an interesting experiment where the attacker ``inadvertently'' uses the correct label as the target label.
In such a condition, backdoor unlearning does not mislead the victim model to unlearning the related memory.
This is because the penalty mechanism of the backdoor unlearning loss (Eq.~\ref{eq_unlearning}) in \sysname can effectively restrain correct memory unlearning.

\section{Conclusion} 
\label{Conclusion}
This paper introduced a backdoor erasing method called \sysname that combined both generative networks and machine unlearning.
\sysname mainly implemented backdoor defense in two steps, namely trigger pattern recovery and trigger pattern unlearning.
In the first step, \sysname discussed about a series of empirical observations about backdoor injection attacks and constrained the trigger pattern recovery problem to be an unknown noise distribution extraction problem.
Then, a entropy maximization generative model was introduced to resolve the problem.
With the recovered trigger patterns, \sysname leveraged them to induce the victim model to reverse the backdoor injection process with a newly designed gradient ascent based machine unlearning method.
Unlike previous works, the novel machine unlearning method got rid of reliance on the access to the training data and was more adaptive to the backdoor erasing scenario.
Extensive experiments were conducted to validate that \sysname could provide stronger defense capability against state-of-the-art backdoor attacks than the existing solutions.

\ifCLASSOPTIONcompsoc
  \section*{Acknowledgments}
\else
  \section*{Acknowledgment}
\fi
This work was supported by the Foundation for Innovative Research Groups of the National Natural Science Foundation of China (No.62121001), the Natural Science Basic Research Program of Shaanxi (2021JC-22), the National Natural Science Foundation of China (61872283, U1764263, No.62072109, No.U1804263), CNKLSTISS, the China 111 Project (No. B16037), Natural Science Foundation of Fujian Province (No. 2021J06013), the Key R\&D Program of Shaanxi Province (No. 2019ZDLGY12-04, No.2020ZDLGY09-06), Innovation Fund of Xidian University (No. YJS2114).

\footnotesize
\bibliographystyle{IEEEtran}
\bibliography{references}

\end{document}